\documentclass[sigconf]{acmart}
\usepackage{multirow}
\usepackage{svg}
\usepackage{soul}
\usepackage{framed}
\usepackage{enumitem}
\newlist{choices}{enumerate}{1}
\setlist[choices,1]{label=\Alph*., leftmargin=*, itemsep=2pt, topsep=2pt}
\usepackage{listings}
\definecolor{lightyellow}{rgb}{1,1,0.7}
\sethlcolor{lightyellow}

\AtBeginDocument{%
  }

\copyrightyear{2026}
\acmYear{2026}
\setcopyright{cc}
\setcctype{by-nc-nd}
\acmConference[CHI '26]{Proceedings of the 2026 CHI Conference on Human Factors in Computing Systems}{April 13--17, 2026}{Barcelona, Spain}
\acmBooktitle{Proceedings of the 2026 CHI Conference on Human Factors in Computing Systems (CHI '26), April 13--17, 2026, Barcelona, Spain}
\acmPrice{}
\acmDOI{10.1145/3772318.3790298}
\acmISBN{979-8-4007-2278-3/2026/04}

\begin{document}

\title{Who You Explain To Matters: Learning by Explaining to Conversational Agents with Different Pedagogical Roles}

\author{Zhengtao Xu}
\email{xuzhengtao@u.nus.edu}
\orcid{0009-0003-4429-4768}
\affiliation{%
  \department{Computer Science}
  \institution{National University of Singapore}
  \city{Singapore}
  \country{Singapore}
}

\author{Junti Zhang}
\email{juntizhang@u.nus.edu}
\orcid{0009-0005-4107-1563}
\affiliation{%
  \department{Institute of Data Science}
  \institution{National University of Singapore}
  \city{Singapore}
  \country{Singapore}
}

\author{Anthony Tang}
\email{tonyt@smu.edu.sg}
\orcid{0000-0003-4293-4082}
\affiliation{%
  \institution{Singapore Management University}
  \city{Singapore}
  \country{Singapore}
}

\author{Yi-Chieh Lee}
\email{yclee@nus.edu.sg}
\orcid{0000-0002-5484-6066}
\affiliation{%
  \department{Computer Science}
  \institution{National University of Singapore}
  \city{Singapore}
  \country{Singapore}
}


\begin{abstract}
  Conversational agents are increasingly used in education for learning support. An application is ``learning by explaining'', where learners explain their understanding to an agent. However, existing research focuses on single roles, leaving it unclear how different pedagogical roles influence learners' interaction patterns, learning outcomes and experiences. We conducted a between-subjects study (N=96) comparing agents with three pedagogical roles (Tutee, Peer, Challenger) and a control condition while learning an economics concept. We found that different pedagogical roles shaped learning dynamics, including interaction patterns and experiences.
Specifically, the Tutee agent elicited the most cognitive investment but led to high pressure. The Peer agent fostered high absorption and interest through collaborative dialogue. The Challenger agent promoted cognitive and metacognitive acts, enhancing critical thinking with moderate pressure. The findings highlight how agent roles shape different learning dynamics, guiding the design of educational agents tailored to specific pedagogical goals and learning phases.
\end{abstract}




\maketitle

\section{Introduction}

Learning by explaining is a pedagogical method in which learners explain their understanding of concepts or processes, either to themselves or to others \cite{ploetzner1999learning}. This approach is widely recognized for enhancing knowledge retention, as it requires learners to actively construct knowledge, organize information, and identify gaps in their own comprehension \cite{chi1994eliciting, mende2024generative}. While self-explanation is beneficial, explaining to a partner is often more effective. A partner’s feedback and questions can immediately expose logical gaps or misconceptions that might otherwise go unnoticed \cite{coleman2014using}.

Conversational agents, widely known as chatbots, are increasingly used in education to provide learners with personalized, on-demand learning support \cite{khosrawi2022conversational, okonkwo2021chatbots}. These agents adopt a variety of pedagogical roles, from tutors providing instruction \cite{lieb2024student} to peers facilitating collaborative dialogue \cite{liu2024peergpt}. A popular application of these agents is the teachable agent, in which a learner takes on the role of a teacher to instruct an AI agent \cite{jin2024teach, rogers2025playing, sabnis2024empowering, wang2025learning, lyu2025exploring}. In this dynamic, the learner teaches concepts to the agent, which in turn asks questions to solidify its ``understanding''. This process, a structured form of learning by explaining, encourages the learner to organize their thoughts and refine their knowledge.

However, the tutor-tutee dynamic is just one of many pedagogical framings that can be used for learning by explaining. Prior work has explored a wide range of agent personas. For instance, researchers have designed agents that act as collaborative peers to facilitate dialogue \cite{liu2024peergpt, lin2024proposed}, and still others have used agents as debaters to provoke critical reflection and argumentative reasoning \cite{favero2024enhancing, tanprasert2024debate}. Each of these designs creates unique interaction dynamics and expectations for the learner.

While these different designs exist, the existing work has primarily focused on a single role at a time and simultaneously varied other factors like implementation and content \cite{wollny2021we, okonkwo2021chatbots}. 
Although insights from these individual studies are valuable, they are challenging to synthesize because differences in task design, learning domains, and agent capabilities often confound the specific effects of the pedagogical role. Consequently, it remains unclear whether observed differences across studies stem from the role itself or from these contextual variations. A direct, controlled comparison is therefore essential to isolate the effect of role design and establish a principled empirical foundation for developing adaptive and role-based conversational agents \cite{gulz2004benefits}. Furthermore, to move beyond simple outcome comparisons, such a controlled setup allows us to systematically trace how distinct roles influence the process of learning---specifically, how they shape the moment-to-moment interaction patterns that ultimately drive those outcomes.
This leaves a gap in our understanding: we lack a systematic comparison of how intentionally designing an agent to adopt a specific pedagogical role influences the entire learning process, from interaction patterns to outcomes and experiences.

To address this gap, our study compares how different pedagogical roles for conversational agents affect the learning-by-explaining process. We designed three role-based agents---``Tutee'', ``Peer'', and ``Challenger''---along with a Control condition. The Tutee agent simulates a novice learner, who asks for clarification and expresses uncertainty. The Peer agent acts as a collaborative partner, offering alternative perspectives and seeking confirmation. Finally, the Challenger agent adopts a Socratic role, questioning the learner’s logic and presenting counterarguments. These distinct roles allow us to systematically measure how these roles influence a learner's interaction patterns, learning outcomes, and experiences.

We conducted a between-subjects study with 96 participants tasked with learning about an undergraduate-level economic concept: ``supply and demand''. Participants first studied the learning material from a textbook, then engaged in five rounds of interaction with their assigned agent to explain a scenario about supply and demand in the U.S. postal service. We analyzed pre-test and post-test scores, survey data on their learning experience, and the conversational logs. We found that different pedagogical roles shaped learning dynamics, including interaction patterns and experiences, while there were no significant differences in objective learning outcomes.
Specifically, the Tutee agent elicited the most cognitive investment from learners but also led to high pressure. The Peer agent fostered high absorption and interest through collaborative dialogue. The Challenger agent promoted cognitive and metacognitive acts, enhancing critical thinking with moderate pressure. The control condition resulted in the poorest learning experience as expected. The findings offer design implications for educational agent designers, emphasizing that these agents should be tailored to specific pedagogical goals and learning phases.

This work makes the following contributions to the fields of human-computer interaction:
\begin{itemize}
    \item We contribute empirical evidence for how different agent roles can be used to shape learning dynamics, including interaction patterns and learning experiences in the context of learning by explaining. This provides designers with a clear understanding of the design trade-offs among different agent roles.
    \item We introduce a theoretically grounded role design by instantiating classic educational interaction types into concrete agent roles, enabling structured investigation of how roles shape learning dynamics.
    \item We provide several design implications for designers of educational agents. The core takeaway is that instead of searching for a single best agent, the goal for designers should be to deploy the right agent for the right pedagogical goal and learning phase.
\end{itemize}

\section{Related Work}

\subsection{Learning by Explaining}
Learning by explaining is the process in which learners deepen their understanding by constructing explanations to themselves or to others \cite{ploetzner1999learning}. When learners explain concepts to themselves (self-explaining), they engage in active knowledge construction by processing new information, putting it into their own words, and making connections to existing knowledge structures \cite{chi1994eliciting, renkl1997learning}. This self-directed process encourages learners to elaborate on provided information, fill in missing steps, and resolve inconsistencies in their understanding \cite{chi2012seeing}.
By contrast, explaining to others transforms the activity into a collaborative and social process, in which participants co-construct knowledge through dialogue, feedback, and shared reasoning \cite{coleman2014using}. This interpersonal dimension introduces additional cognitive demands, as explainers must consider the perspective of their audience, adapt their communication to be comprehensible, and respond to questions or challenges that create cognitive conflict \cite{piaget1977development} or reveal gaps in their own understanding \cite{bargh1980cognitive}. 
Both forms of learning by explaining are grounded in constructivist learning theory, which emphasizes the active role of learners in building their own knowledge through meaningful engagement with content \cite{palincsar1998social}.

Prior studies have shown that learning by explaining offers significant cognitive benefits for learners. \citeauthor{chi1989self} identified the ``self-explanation effect'', where learners who are more successful at problem-solving tend to generate more self-explanations while studying \cite{chi1989self}. Explaining to others brings additional benefits through the social aspect of learning. \citeauthor{webb2003promoting} found that when students explain things to their peers, they not only help their partners learn, but they also solidify and improve their own understanding through the articulation process \cite{webb2003promoting}. In addition, explaining provides opportunities for learners to engage in generative cognitive processes, including organizing information, making connections between concepts, and identifying gaps in their knowledge \cite{fiorella2016eight}, while simultaneously exercising metacognitive control through self-monitoring and planning \cite{flavell1979metacognition, zimmerman2002becoming}. Both of these processes promote procedural flexibility and conceptual change because they require learners to actively construct coherent mental models rather than passively receive information \cite{rittle2006promoting}.

Despite its potential benefits, learning by explaining faces many limitations and challenges. One big concern is that learners may generate poor or incorrect explanations, which reinforce misconceptions rather than promote learning \cite{chi2001learning}. Additionally, learners often lack explainees, particularly in individual learning contexts where peers or instructors may not be available \cite{webb2003promoting}. These challenges highlight the potential of virtual agents as explanation recipients, as they can provide consistent availability and adaptive scaffolding—such as corrective feedback or guiding questions—that help learners detect and rectify potential misconceptions, thereby improving explanation quality \cite{kuhail2023interacting}. 
Given these benefits, we use conversational agents as explanation partners to support learning by explaining and build on previous findings about their potential advantages.

\subsection{Conversational Agents in Education}

Conversational agents have increasingly been explored as educational tools with the development of AI \cite{khosrawi2022conversational, kuhail2023interacting}. These systems provide unique opportunities for learners to engage with educational content more flexibly, such as enabling interaction at one's own pace and outside the constraints of traditional classroom settings \cite{weber2021pedagogical}. The conversational nature of these agents can create low-stakes environments with psychological safety \cite{edmondson1999psychological} where learners feel more comfortable asking questions, making mistakes, and exploring concepts without fear of judgment  \cite{cottrell1972social, hawanti2023ai}. Moreover, conversational agents can provide repeated opportunities for practice and feedback without the limitations of human availability while maintaining relatively consistent educational support.

Educational conversational agents can be classified into several types based on their pedagogical goals and interaction designs. The most common category is \textit{teaching agents}: these serve as virtual tutors that provide direct instruction and personalized feedback \cite{kulik2016effectiveness}. For example, \citeauthor{lieb2024student} introduced NewtBot, a physics education chatbot that functions as a personalized automated tutor by guiding students step by step through problem-solving tasks and providing corrective feedback to support secondary students in physics learning \cite{lieb2024student}. Another major category is \textit{peer agents}, specifically designed to facilitate collaborative dialogues and promote peer-to-peer interactions among learners \cite{kuhail2023interacting}. For instance, \citeauthor{liu2024peergpt} explored how a peer agent's role as either a team moderator or participant affects peer conversations in children's collaborative learning, identifying potential design improvements for effective peer agents \cite{liu2024peergpt}. Pertinent to our work, \textit{teachable agents} adopt a learning-by-teaching approach. In this approach, the agent acts as a novice and asks learners to guide them along a learning route. This setup prompts learners to articulate and demonstrate their understanding by explaining concepts and procedures to the agent, thereby strengthening their own comprehension through the act of teaching. Recent studies have further demonstrated the effectiveness of teachable agents in the context of LLM-based systems, where learners teach the agent in domains such as programming \cite{jin2024teach, rogers2025playing}, mathematics \cite{sabnis2024empowering}, and language learning \cite{wang2025learning}. Lastly, rather than focusing primarily on direct knowledge transmission, \textit{motivational agents} aim specifically to enhance learner engagement, persistence, and emotional well-being through affective support and behavioral change techniques. For example, \citeauthor{ceha2021can} demonstrated that conversational agents using affiliative humor could significantly improve learners' motivation, effort, and emotional experiences during a teaching task \cite{ceha2021can}. 

Unlike teaching or motivational agents, both peer and teachable agents fundamentally rely on explanation-based interactions, where learners construct knowledge by explaining their understanding to the agent. While prior work has established the general value of such explanation relationships, it often treats ``learning by explaining'' as a single, uniform activity. What remains less understood is how to intentionally shape the nature of this explanatory process to achieve more specific pedagogical goals. Our work addresses this gap by positing that the agent's pedagogical role is the key design element for this purpose. We therefore explore how agents designed with different roles can be used to guide learners through different cognitive processes and experiences, tailored to specific learning goals.

\subsection{Role-based Conversational Agents in Learning by Explaining}

Learning by explaining often relies on a partner who structures the interaction and elicits articulation \cite{holmes2007designing}. The role dynamics of these learning partners influence the interaction patterns in explanation tasks, such as the level of detail and the focus of the explanation \cite{bargh1980cognitive}. 
From a social cognitive perspective \cite{bandura1986social}, the roles of a partner matter because they act as social cues that shape learners' attention, effort, and self-evaluation. 
For example, a collaborative and non-judgmental partner can foster a high degree of psychological safety, encouraging learners to take risks and share ideas without fear \cite{edmondson1999psychological}. Conversely, a partner designed to challenge a learner's statements can induce cognitive conflict, a state of mental discomfort that can motivate learners to re-evaluate their understanding and achieve deeper conceptual change \cite{piaget1977development}.
Therefore, for conversational agents used in educational settings, the agent’s role is a crucial design element that can systematically influence explanatory behavior and the learning process.

To investigate these dynamics, our work defines three agent roles by combining \citeauthor{moore1989three}'s foundational framework of educational interaction \cite{moore1989three} with established learning theories. \citeauthor{moore1989three} identifies three fundamental types of interaction in learning: learner-instructor, learner-learner, and learner-content. We instantiate these categories through roles that capture well-documented pedagogical traditions.
The \textbf{Tutee agent} simulates a novice, positioning the user as instructor in a reversed learner-instructor dynamic; this draws on research on the protégé effect, which shows that teaching enhances the explainer's learning \cite{chase2009teachable}. The \textbf{Peer agent} represents a collaborative learner–learner interaction, reflecting theories of peer learning and socially shared metacognition, where equal partners foster psychological safety and co-construction \cite{vauras2004socially}. The \textbf{Challenger agent} embodies an expert instructor in a traditional learner-instructor interaction, adopting a Socratic style \cite{paul2019thinker} and grounded in work on cognitive conflict and conceptual change, where critical questioning deepens reasoning \cite{piaget1977development}. Finally, our \textbf{Control condition} serves as a baseline, representing a pure learner–content interaction. By grounding our roles in \citeauthor{moore1989three}'s framework and these complementary educational theories, we can systematically compare how distinct pedagogical framings influence learners' interaction patterns (RQ1), learning outcomes, and learning experiences (RQ2).

\section{Research Questions}

Our research first aims to explore how different agent roles in learning-by-explaining contexts influence users' interaction patterns. We conceptualize these patterns, such as the types of statements learners make in their explanations, as manifestations of their cognitive and metacognitive processes. Understanding how an agent's role can shape these processes is the first step toward aligning its design with a specific learning objective. 

Second, we aim to examine the effects of these different role-based interactions on learners' learning outcomes and experiences, and explore the underlying reasons for these effects. By analyzing these effects, we can gain a deeper understanding of not just the impact of different interaction patterns on the learning process, but more importantly, how these observable behaviors are ultimately perceived and experienced by the learner.

Accordingly, we seek to answer the following two research questions:
\begin{itemize}
    \item \textbf{RQ1:} \textit{How do different conversational agent roles (Tutee, Peer, Challenger, Control) influence how learners engage and explain topics (interaction patterns) in learning-by-explaining contexts?}
    \item \textbf{RQ2:} \textit{How do different conversational agent roles (Tutee, Peer, Challenger, Control) influence learners' learning outcomes and learning experiences?}
\end{itemize}

\section{Methods}

\subsection{Conversational Agents Design}

We designed four conversational agent conditions (Tutee, Peer, Challenger, Control), each representing distinct pedagogical roles and interaction strategies to systematically explore their effects on learning by explaining. All agents were powered by OpenAI’s GPT-4o, which handled both message analysis and response generation. All prompts used are provided in Appendix~\ref{appendix:prompts}.

To standardize the start of the interaction, each agent initiated the first turn with a role-consistent prompt:
\begin{itemize}
    \item \textbf{Tutee:} \textit{``Could you please remind me of the basic definitions of 'demand', 'supply', 'price', and 'equilibrium'?''}
    \item \textbf{Peer:} \textit{``I'm really curious about your process—can you share your first step to solving the problem?''}
    \item \textbf{Challenger:} \textit{``Could you please provide a clear, step-by-step approach for solving this problem?''}
    \item \textbf{Control:} \textit{``Please share your approach to solving the problem.''}
\end{itemize}

\begin{figure*}[htbp]
  \centering
  \includegraphics[width=0.95\linewidth]{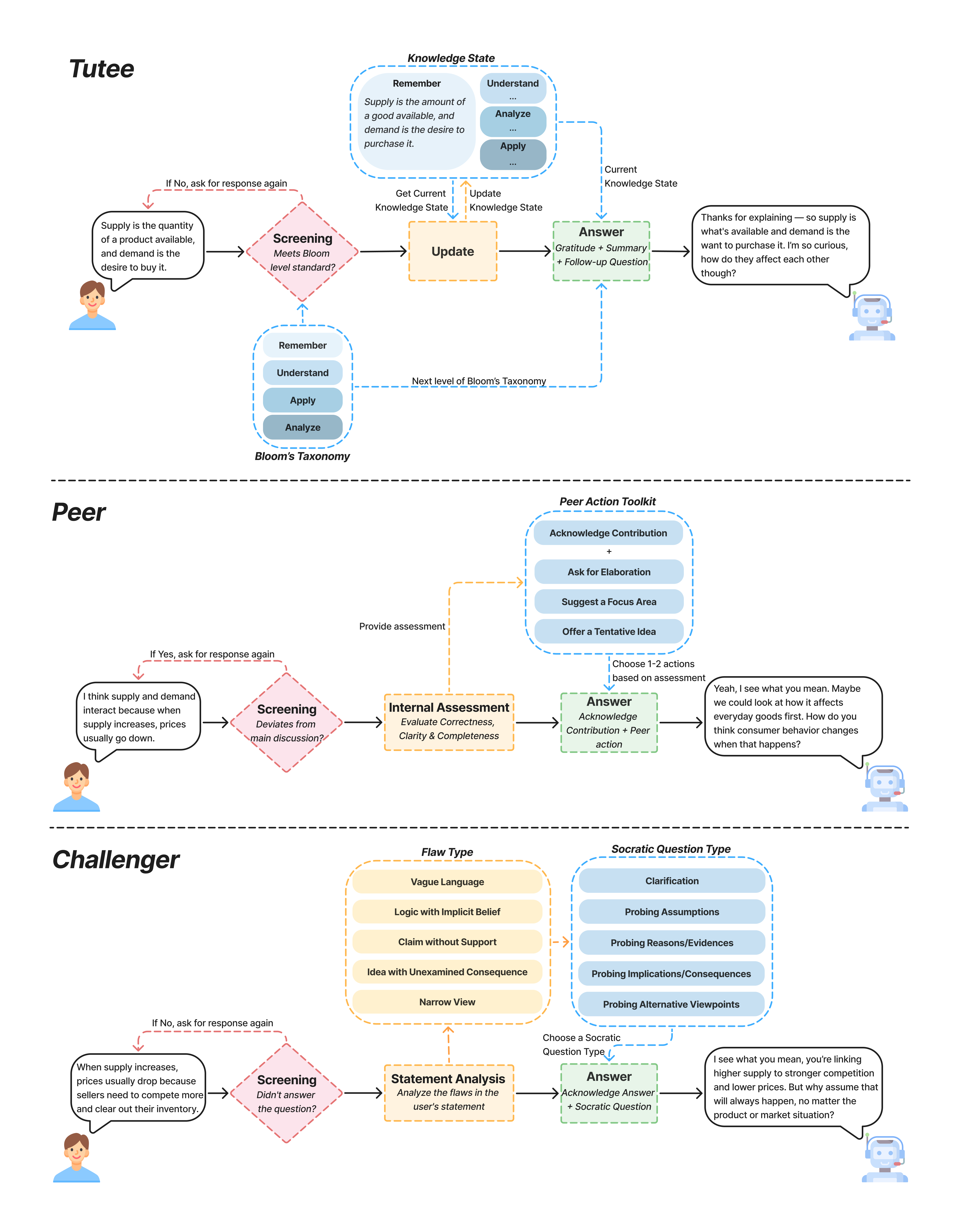}
  \caption{Workflows of the three agent roles (Tutee, Peer, and Challenger). The figure illustrates the decision flow and key actions for each role. Note that the agent responses shown in the figure are only schematic examples of the response structure; in our study, the agent responses under each condition were constrained to around 80 words.}
  \label{fig:agents_workflows}
  \Description{The figure presents workflows of the three agent roles: Tutee, Peer, and Challenger. In the Tutee condition, the agent screens the participant’s input against Bloom’s taxonomy levels. The flow updates the agent’s knowledge state and responds with gratitude, summaries, and follow-up questions, modeling a learner role. In the Peer condition, the agent conducts an internal assessment of the participant’s contribution for correctness and clarity, then provides peer-style actions such as acknowledging contributions, asking for elaboration, or offering tentative ideas. In the Challenger condition, the agent screens input for vague or unsupported statements, performs statement analysis to identify flaws, and then responds with Socratic questions probing assumptions, reasoning, implications, or alternative viewpoints. Example messages are shown in speech bubbles, but actual experimental responses were constrained to about 80 words. The three workflows illustrate contrasting pedagogical strategies: absorption and engagement through teaching (Tutee), collaborative exchange (Peer), and stimulating critical reflection (Challenger).}
\end{figure*}

Following this standardized entry point, each agent proceeded with a distinct workflow. Figure~\ref{fig:agents_workflows} illustrates the workflows of the Tutee, Peer, and Challenger agents, and the subsections below provide detailed descriptions of how each agent processed learner input and generated responses. An example interaction transcript for each role condition is provided in Appendix~\ref{appendix:transcripts}.

\subsubsection{Tutee agent}
The Tutee agent simulated a novice student progressing through Bloom’s taxonomy of learning objectives~\cite{anderson2001taxonomy}. User inputs were first screened against Bloom-level standards. Valid responses were integrated into the agent’s knowledge state, which advanced sequentially from remember to understand, apply, and analyze. The agent then produced an output that combined gratitude, a brief summary of what had been ``learned'', and a follow-up question in the voice of a novice (e.g., \textit{``Thanks for explaining — so supply is what’s available and demand is the want to purchase it. I’m so curious, how do they affect each other though?''}). This design ensured the agent remained authentic as a tutee rather than displaying full expert knowledge.

\subsubsection{Peer agent}
The Peer agent acted as a collaborative partner based on peer feedback frameworks~\cite{gielen2010improving}. After screening whether the user’s response was on-topic, the agent internally assessed its correctness, clarity, and completeness. It then acknowledged the contribution and flexibly selected one or two peer actions (ask for elaboration, suggest a focus area, offer a tentative idea) from the framework. For example: \textit{``Yeah, I see what you mean. Maybe we could look at how it affects everyday goods first. How do you think consumer behavior changes when that happens?''} This approach aimed to maintain a supportive and co-constructive dialogue.

\subsubsection{Challenger agent}
The Challenger agent adopted a Socratic questioning approach~\cite{paul2019thinker} to critically probe participants’ reasoning. User inputs were first screened for adequacy, followed by a statement analysis that identified flaws such as vague language. Based on this analysis, the agent selected an appropriate Socratic question type (clarification, probing assumptions, probing reasons and evidence, probing implications and consequences, probing alternative viewpoints). Each response began with a brief acknowledgment and then challenged the underlying reasoning. For example: \textit{``I see what you mean, you’re linking higher supply to stronger competition and lower prices. But why assume that will always happen, no matter the product or market situation?''}

\subsubsection{Control condition}
The Control condition minimized interaction to simulate a pure self-explaining scenario. The agent produced only brief, generic acknowledgments such as \textit{``Noted. Please review and consider if there are any additional points you wish to include.''} without embodying an interactive role or contributing substantive feedback. To avoid repetition, the control agent used a small set of neutral templates for brief acknowledgments, ensuring varied wording across rounds while maintaining functional equivalence.

\subsection{Learning Topic and Task}
\label{sec:learning_topic}
Our study was designed for an undergraduate-level learning context, where learners are expected to have the cognitive maturity to engage in abstract reasoning, critical reflection, and structured explanation. Guided by this scope, we established three criteria for selecting the learning topic: (1) it should involve conceptual understanding rather than rote recall, (2) it should not constitute common prior knowledge, and (3) it should be sufficiently learnable within a short session to allow meaningful explanations. Based on these criteria, we chose ``supply and demand'' from introductory economics, selecting Sections 3.1–3.3 of OpenStax \emph{Principles of Economics 3e}\footnote{\url{https://openstax.org/books/principles-economics-3e/pages/3-introduction-to-demand-and-supply}}. The materials cover the definitions of demand, supply, and equilibrium; factors that shift demand and supply; and the systematic four-step process for analyzing changes in equilibrium price and quantity. We condensed the content to fit a 15-minute study window while retaining sufficient conceptual depth for the subsequent task.

\begin{figure*}[htbp]
  \centering
  \includegraphics[width=\linewidth]{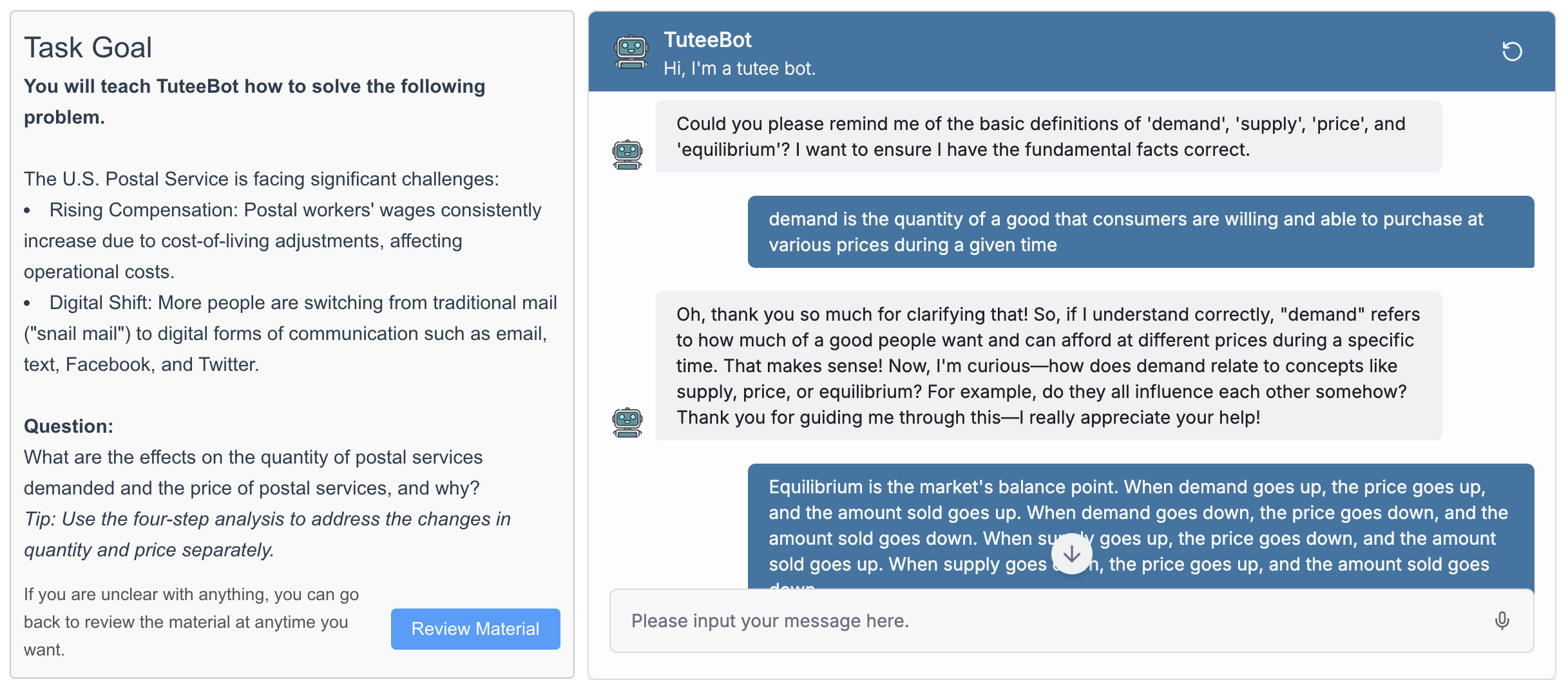}
  \caption{Interaction interface used in the study under the Tutee condition. The left panel displays the task problem, while the right panel shows the chat interface where participants explained to the agent.}
  \label{fig:chatbot-ui}
  \Description{The figure shows the user interface under the Tutee condition. On the left, a task panel outlines the problem context about challenges faced by the U.S. Postal Service, including rising compensation and a digital shift, followed by a guiding question. On the right, the chat interface displays the conversation between the participant and TuteeBot. The bot asks for definitions of key economics terms such as ``demand'' and ``equilibrium'', and the participant provides explanations in response. The interface includes message bubbles with alternating colors for clarity, a scrolling area for past exchanges, and a text input box at the bottom where participants type their answers. This design emphasizes the teaching interaction, where participants act as instructors guiding the bot.}
\end{figure*}

To ground the learning-by-explaining task, we used a scenario-based problem on the U.S. Postal Service highlighting two trends: (1) rising compensation for postal workers due to cost-of-living adjustments, and (2) a digital shift away from traditional mail toward electronic communication. Participants applied supply-and-demand concepts to explain the expected effects on the quantity and price of postal services. The full problem text is provided in Appendix~\ref{appendix:task}. Figure~\ref{fig:chatbot-ui} displays the system interface where participants explained the problem to the Tutee agent.

\subsection{Study Procedure}

\begin{figure*}[tbp]
  \centering
  \includegraphics[width=\linewidth]{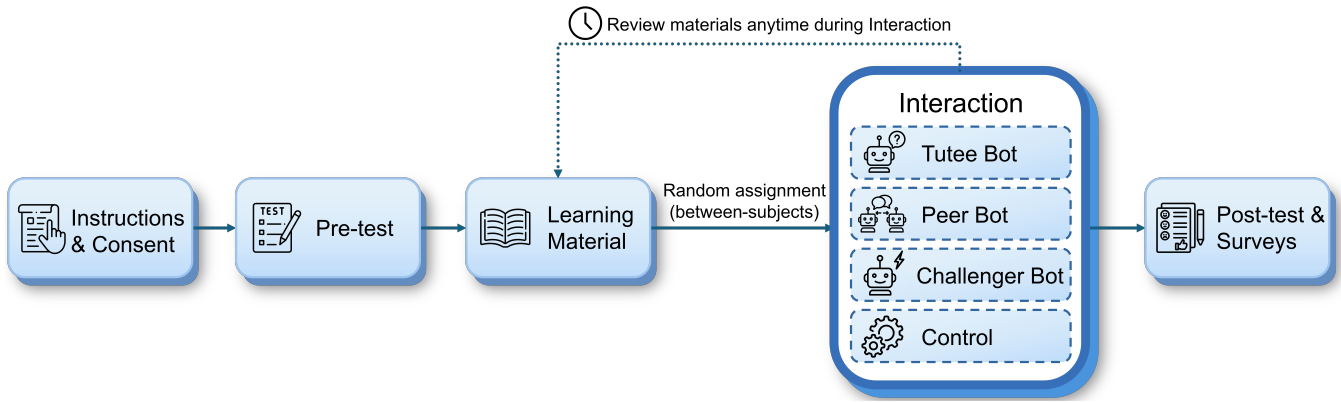} 
  \caption{Experimental procedure. Participants read instructions and gave consent, completed a pre-test (5 min), studied learning materials (15 min), engaged in five rounds of interaction with agents (20 min), and finished with a post-test and surveys (20 min).}
  \label{fig:procedure}
  \Description{The figure presents the experimental procedure in a sequential flowchart. Participants first read instructions and provided consent. They then completed a pre-test lasting 5 minutes, followed by 15 minutes of studying the learning materials. After this, they engaged in five rounds of interaction with conversational agents over 20 minutes. Finally, participants completed a 20-minute post-test and a set of surveys. Each stage is visually represented as a separate step in the flow, emphasizing the structured order and time allocation of the study.}
\end{figure*}

The experiment followed a structured procedure (see Figure~\ref{fig:procedure}) lasting approximately 60 minutes. After entering the experiment, participants first read instructions explaining the study procedures and provided informed consent. Participants then completed a pre-test (5 minutes) to assess their prior knowledge of supply and demand concepts. Next, during the learning phase (15 minutes), participants studied standardized materials covering fundamental economic principles of supply and demand. Following the learning phase, participants were randomly assigned to a condition and explicitly instructed on their agent's pedagogical role (e.g., ``You will teach TuteeBot'') and the task goal. They then entered the interaction phase (20 minutes), where they engaged in five rounds of learning-by-explaining activities with their assigned agent. In each round, they applied the learned concepts to explain a microeconomic scenario: the U.S. Postal Service case. During this phase, participants could return to review the learning materials if needed. Finally, participants completed a post-test and surveys assessing their learning experience (20 minutes).

\subsection{Participants}

We recruited 96 participants through CloudResearch's Connect platform\footnote{\url{https://connect.cloudresearch.com/}}, a crowdsourcing platform designed for online research. All participants were located in the United States and fluent in English. Participants ranged in age from 18 to 73 years ($M = 37.02, SD = 12.25$), with a gender distribution of 55 males and 41 females. Regarding racial and ethnic background, 56 participants identified as White, 22 as Black or African American, 11 as Asian (including Chinese, Indian, Filipino, Vietnamese, and Korean), and 7 preferred not to say. For educational background, 14 participants had completed high school, 27 had some college experience (including associate degrees), 35 held bachelor's degrees, 18 had graduate-level education (master's, doctoral, or professional degrees), and 2 preferred not to say.

Participants were randomly assigned to one of four conditions ($n=24$ per group). Pre-test scores confirmed that participants were not entering the study with deep prior knowledge, with a moderate average of $M=12.67 (SD=3.77)$ out of 20. A one-way ANOVA also found no significant differences across conditions ($F(3,92)=0.73, p=.54$), indicating that the groups were comparable.
Each participant received 10 USD for the one-hour study.

\subsection{Measurements}
\label{sec:measurements}

To comprehensively evaluate the learning-by-explaining process, we employed a mixed-methods approach involving behavioral logs, qualitative coding, objective knowledge tests, and self-reported surveys. Table~\ref{tab:constructs_overview} provides a comprehensive overview of all constructs and their corresponding measurement methods used in this study.

\begin{table*}[t]
\centering
\small
\renewcommand{\arraystretch}{1.20}
\setlength{\tabcolsep}{4pt}
\caption{Summary of all constructs and measurement methods in the study.}
\label{tab:constructs_overview}
\begin{tabular}{p{3.2cm} p{4.0cm} p{9.0cm}}
\toprule
\textbf{Construct Category} & \textbf{Construct / Variable} & \textbf{Measurement Method} \\
\midrule

\multicolumn{3}{l}{\textbf{1. Behavioral Log Data}} \\
\midrule
Behavioral Activity 
& Word Count 
& Total number of words sent by the participants across all five explanation rounds. \\

& Frequency of Reviewing Material
& Number of times a participant scrolled back to the learning material after the initial study phase. \\

& Duration of Reviewing Material 
& Total time spent viewing the material after the initial study phase. \\

\midrule
\multicolumn{3}{l}{\textbf{2. Interaction Patterns (Qualitative Coding)}} \\
\midrule

Cognitive Acts 
& (a) Stating Facts/Definitions \newline (b) Elaborating Causality \newline (c) Providing Examples \newline (d) Comparing/Contrasting
& Messages coded for cognitive processing. \\

Metacognitive Acts 
& (a) Self-Monitoring \newline (b) Planning \newline (c) Self-Correction \newline (d) Seeking Confirmation
& Messages coded for metacognitive reflection or regulation of thinking. \\

Interaction Acts 
& (a) Discourse Coordination \newline (b) Feedback \& Evaluation 
& Messages coded for conversational management or evaluative responses. \\

\midrule
\multicolumn{3}{l}{\textbf{3. Knowledge Test}} \\
\midrule

Objective Learning Outcome  
& Pre-test \& Post-test Scores 
& 20-item multiple-choice test. The difference between pre- and post-test represents the learning outcome. \\

\midrule
\multicolumn{3}{l}{\textbf{4. Self-Reported Measures}} \\
\midrule

Intrinsic Motivation (IMI) 
& Perceived Competence \newline Pressure/Tension \newline Interest/Enjoyment 
& 7-point Likert scale adapted from the IMI~\cite{mcauley1989psychometric}. \\

Flow State (FSS) 
& Learning Absorption \newline Learning Fluency 
& 7-point Likert scale adapted from the FSS~\cite{peifer2021advances}. \\

Critical Thinking 
& Perceived Critical Thinking Score 
& 7-point Likert scale adapted from~\citeauthor{tanprasert2024debate}~\cite{tanprasert2024debate}. \\

\bottomrule
\end{tabular}
\Description{The table summarizes all constructs and measurement methods used in the study across four categories. 
Category 1, Behavioral Log Data, includes word count, frequency of reviewing material, and duration of reviewing material, each defined as quantitative logs of participants' engagement with the agent and learning materials. 
Category 2, Interaction Patterns (Qualitative Coding), lists cognitive acts (stating facts or definitions, elaborating causality or principles, providing examples or analogies, comparing or contrasting), metacognitive acts (self-monitoring, planning, self-correction, seeking feedback or confirmation), and interaction acts (discourse coordination and feedback and evaluation). 
Category 3, Knowledge Test, describes objective learning outcome using pre-test and post-test scores from a 20-item multiple-choice test, with learning outcome calculated as the difference between post- and pre-test scores. 
Category 4, Self-Reported Measures, covers intrinsic motivation subscales (perceived competence, pressure or tension, interest or enjoyment), flow state measures (learning absorption and learning fluency), and perceived critical thinking, each measured with 7-point Likert scales adapted from established instruments.}
\end{table*}

\subsubsection{Behavioral log data}
We logged quantitative metrics regarding users' engagement with the agent and materials. As detailed in Table~\ref{tab:constructs_overview}, these metrics included \textit{Word Count} (engagement with the agent) as well as the \textit{Frequency} and \textit{Duration of Reviewing Material} (engagement with the materials).

\subsubsection{Chat message analysis}
To analyze the interaction patterns, we conducted a qualitative analysis of the user messages in five interaction rounds. We developed a coding scheme to categorize user messages into cognitive, metacognitive, and interaction acts (see Table~\ref{tab:qualitative_coding_scheme} for the full scheme). The detailed procedure regarding the scheme's development is described in \ref{sec:qual}.

\begin{table*}[htbp]
\small
\centering
\renewcommand{\arraystretch}{1.10}
\setlength{\tabcolsep}{2pt}
\caption{Final coding scheme used to analyze participants’ chat messages during the learning-by-explaining task. Each category is defined and illustrated with an example user message.}
\label{tab:qualitative_coding_scheme}
\begin{tabular}{p{2.8cm} p{3.6cm} p{4.4cm} p{6.2cm}}
\toprule
\textbf{Category} & \textbf{Subcategory} & \textbf{Definition} & \textbf{Example of User Message} \\
\midrule
\multicolumn{4}{l}{\textbf{Dimension 1: Cognitive \& Metacognitive Actions}} \\
\midrule
\multirow{5}{*}{\textit{Cognitive Acts}} 
& Stating Fact/Definition 
& Stating factual information or definitions of concepts. 
& \textit{Demand is about how much of something people want to buy and are actually able to pay for.} ($P2_T$) \\
\cmidrule(lr){2-4}
& Elaborating Causality/Principle 
& Explaining causal relationships or principles involving variables. 
& \textit{As the demand for postal services decreases due to people using email, the cost to provide services increases. This increase in cost is due to the increase of workers' wages and increased operational costs.} ($P5_C$)\\
\cmidrule(lr){2-4}
& Providing Example/Analogy 
& Giving an example or analogy to clarify a concept. 
& \textit{For example, a certain percent off if they buy postage online, like companies such as eBay offer.} ($P6_C$)\\
\cmidrule(lr){2-4}
& Comparing/Contrasting 
& Highlighting differences or similarities between entities. 
& \textit{If demand increases, price increases and equilibrium quantity increases. If the supply decreases, then the price also increases, but the equilibrium quantity decreases.} ($P14_T$)\\
\midrule
\multirow{4}{*}{\textit{Metacognitive Acts}} 
& Self-Monitoring 
& Expressing awareness of one’s own cognitive state. 
& \textit{Absolutely, I was just thinking that. I'm not sure how post offices price things.} ($P11_P$)\\
\cmidrule(lr){2-4}
& Planning/Strategizing 
& Outlining a plan or strategy for problem solving. 
& \textit{First, let's identify the starting equilibrium price and quantity...} ($P4_C$) \\
\cmidrule(lr){2-4}
& Self-Correction 
& Correcting a previous statement. 
& \textit{Yes, thank you. I meant to say supply. So I shift the supply curve left.} ($P19_C$) \\
\cmidrule(lr){2-4}
& Seeking Feedback/Confirmation 
& Asking to confirm understanding or receive feedback. 
& \textit{I think it would make the price increase, right?} ($P20_P$) \\
\midrule
\multicolumn{4}{l}{\textbf{Dimension 2: Interaction Actions}} \\
\midrule
\multirow{2}{*}{\textit{Discourse Coordination}} 
& Requesting Task Clarification 
& Asking for clarification about the task requirements. 
& \textit{Can you restate the problem?} ($P11_C$)\\
\cmidrule(lr){2-4}
& Turn Closing/Refusal to Elaborate 
& Indicating the end of a turn or declining further elaboration. 
& \textit{Nothing else comes to my mind. ($P17_B$)} \\
\midrule
\multirow{2}{*}{\textit{Feedback \& Evaluation}} 
& Agreement/Affirmation 
& Expressing agreement or affirmation with a statement. 
& \textit{Yes, your understanding is correct.} ($P5_T$) \\
\cmidrule(lr){2-4}
& Disagreement/Rebuttal 
& Expressing disagreement or giving a counterargument. 
& \textit{No, I think this is a societal shift that basic economics won't solve.} ($P15_C$) \\
\midrule
\textit{Other} 
& Off-Topic Chat 
& Producing content unrelated to the task. 
& \textit{I can't read the box.} ($P6_P$)\\
\bottomrule
\end{tabular}
\Description{The table shows the final coding scheme used to analyze participants’ chat messages during the learning-by-explaining task. It is organized into two main dimensions plus an additional category.
Dimension 1: Cognitive and Metacognitive Actions includes:
– Cognitive Acts: stating facts or definitions, elaborating causal relationships, providing examples or analogies, and comparing or contrasting.
– Metacognitive Acts: self-monitoring (awareness of one’s cognitive state), planning or strategizing, self-correction of earlier statements, and seeking feedback or confirmation.
Dimension 2: Interaction Actions includes:
– Discourse Coordination: requesting task clarification and turn closing/refusal to elaborate.
– Feedback and Evaluation: expressing agreement or affirmation, or expressing disagreement and rebuttal.
Other: off-topic chat, when participants produced content unrelated to the task.
Each category is defined and illustrated with a representative user message, such as: ``Demand is about how much of something people want to buy…'' (Stating Fact/Definition), or ``I think it would make the price increase, right?'' (Seeking Feedback/Confirmation).}
\end{table*}

\subsubsection{Knowledge test scores}
To objectively measure learning outcomes, participants were asked to complete a knowledge test before (pre-test) and after (post-test) the interaction. The test consisted of 20 multiple-choice questions assessing conceptual understanding and application of the topic, adapted from the OpenStax textbook mentioned in Section~\ref{sec:learning_topic}. Each question included one correct answer, three incorrect answers, and an ``I don't know'' option to mitigate guessing~\cite{muijtjens1999effect}. The complete set of test questions is provided in Appendix~\ref{appendix:test}.

\subsubsection{Self-reported measures}
After the interaction, participants completed a survey with several scales. All items were rated on a 7-point Likert scale.

To measure affective experience, we used three task-relevant subscales from the Intrinsic Motivation Inventory (IMI)~\cite{mcauley1989psychometric}: \textit{Perceived Competence}, \textit{Pressure/Tension} and \textit{Interest/Enjoyment} (1 = strongly disagree, 7 = strongly agree). We reviewed all items and adjusted their wording to fit our study context. All three subscales showed high internal consistency (\textit{Perceived Competence}: $\alpha = .922$, \textit{Pressure/Tension}: $\alpha = .919$, \textit{Interest/Enjoyment}: $\alpha = .932$).

We measured participants' flow state using two dimensions from the Flow Short Scale (FSS)~\cite{peifer2021advances}, following the adaptation by~\citeauthor{chung2024enhancing}~\cite{chung2024enhancing}: \textit{Learning Absorption} (the flow state of deep immersion and total concentration in the learning task) and \textit{Learning Fluency} (the perceived smoothness and continuity of the mental process during the learning task) (1 = not at all, 7 = very much). We took the mean score of the six questions (Q2, Q4, Q5, Q7, Q8, Q9) as the participants’ learning fluency. For learning absorption, we retained three items (Q1, Q3, Q6) and dropped one item (Q10) due to ambiguity in reading. Both subscales showed acceptable internal consistency (\textit{Learning Absorption}: $\alpha = .708$, \textit{Learning Fluency}: $\alpha = .882$).

Self-reported critical thinking was measured using a scale adapted from the work of~\citeauthor{tanprasert2024debate}~\cite{tanprasert2024debate}. The scale is structured around six types of critical thinking (1 = strongly disagree, 7 = strongly agree). We calculated an overall critical thinking score by averaging all items across the six types. The overall scale demonstrated high internal consistency in our sample ($\alpha = .963$).

\subsubsection{Open-ended questions}
The post-study survey concluded with three open-ended questions. These questions prompted participants to provide qualitative feedback on the perceived impact of the interaction on learning, affective experience during the task, and additional feedback. The questions were:
\begin{enumerate}[label=\arabic*.]
    \item \textit{Do you think interacting with the bot helped your learning and problem-solving process? If so, how did it help?}
    \item \textit{Please describe your feelings and experiences when interacting with the bot.}
    \item \textit{Do you have any additional feedback from your interaction with the bot?}
\end{enumerate}

\subsection{Data Analysis}

\subsubsection{Quantitative analysis}
For the quantitative analyses, statistical tests were selected based on the characteristics of the data. 
When the data met both normality and homogeneity of variance assumptions, we conducted one-way Analyses of Variance (ANOVA) with Tukey's post-hoc tests. 
When normality was met but the assumption of homogeneity of variances was violated, Welch’s ANOVA was used with Games–Howell post-hoc tests. 
For non-normally distributed data, we used Kruskal–Wallis H tests, followed by Dunn’s post-hoc tests with Bonferroni corrections. 
Statistical significance was set at an alpha level of .05.

\subsubsection{Qualitative analysis}
\label{sec:qual}

We conducted two qualitative analyses to better understand participants’ interactions with the agent and their perceptions of the learning-by-explaining task.

\definecolor{myagree}{HTML}{42BC9A}
\definecolor{mycausality}{HTML}{FCBD67}  
\definecolor{myexample}{HTML}{79ADD6}

\providecommand{\codeAgree}[1]{{\sethlcolor{myagree!60}\hl{#1}}}
\providecommand{\codeElab}[1]{{\sethlcolor{mycausality!60}\hl{#1}}}
\providecommand{\codeExample}[1]{{\sethlcolor{myexample!60}\hl{#1}}}

The first analysis focused on the chat messages participants wrote during the learning-by-explaining task. To systematically analyze these interactions, we developed a coding scheme using a deductive-inductive approach derived from established theoretical frameworks on explanatory dialogue \cite{chi1994eliciting}, metacognition \cite{flavell1979metacognition}, and interaction analysis \cite{de2006content}. Two researchers collaboratively engaged in an iterative process to develop the scheme. An initial set of codes was defined based on the literature, followed by pilot coding on a subset of data to test applicability. Through discussion, the researchers refined category definitions and resolved ambiguities to finalize the scheme presented in Table~\ref{tab:qualitative_coding_scheme}. Each chat message was treated as a single unit of analysis and could be assigned multiple codes if it contained more than one action. For example, the following message received three codes (\colorbox{myagree!60}{\phantom{M}}: Agreement/Affirmation; \colorbox{mycausality!60}{\phantom{M}}: Elaborating Causality/Principle, \colorbox{myexample!60}{\phantom{M}}: Providing Example/Analogy).

\begin{quote}
\codeAgree{Yes}, \codeElab{the postal office would need to reduce costs in other ways or find more services to sell to consumers}. \codeExample{For example, high quality printing or selling stamps, envelopes and boxes at a higher-than-equilibrium rate due to convenience.} ($P17_P$)
\end{quote}

The second analysis focused on participants’ responses to open-ended questions at the end of the study, following a thematic analysis approach \cite{braun2006using}. Two researchers independently read through all responses and generated initial codes. These codes were then discussed, compared, and refined to develop broader themes.

For both analyses, two researchers independently coded a 25\% subset of the data, achieving high inter-rater reliability (Krippendorff’s alpha = 0.831 for the chat message coding; 0.875 for the open-ended questions coding). All disagreements were resolved through discussion, after which the finalized coding frameworks were applied to the remaining data by one researcher.

\section{Results}
\label{sec:results}
This section presents the primary findings of our study, organized according to our two research questions. We first address RQ1 by analyzing and comparing the differences in user interaction patterns across the four conditions. Following this, we address RQ2 by evaluating the effects of these different conversational agent roles on learners' learning outcomes and their learning experiences. Full statistical results are provided in Appendix~\ref{appendix:stats}.

In the presentation of qualitative data, participants are denoted using the format $Px_y$, where $x$ is the participant number, and $y$ is the experimental condition: $T$ for the Tutee group, $P$ for the Peer group, $C$ for the Challenger group, and $B$ for the Baseline (Control) group.

\subsection{RQ1 - Differences in User Interaction Patterns}

\subsubsection{Quantitative analysis of behavioral logs}

\begin{figure*}[htbp]
  \centering
  \includegraphics[width=\textwidth]{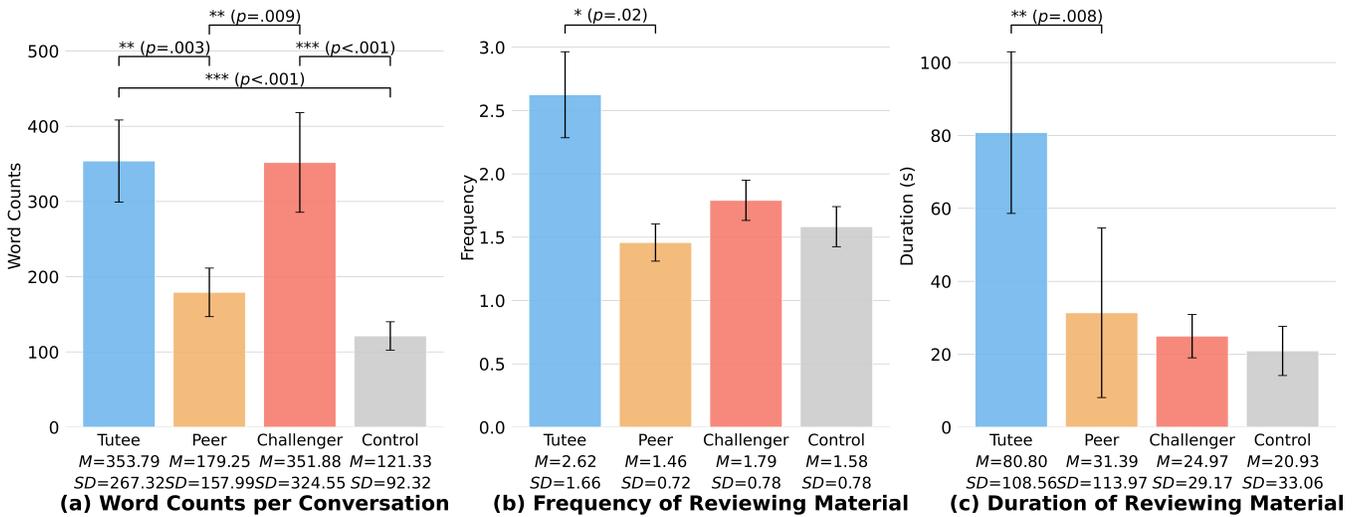}
  \caption{User interaction metrics across agent roles. 
  (a) Word counts per conversation; 
  (b) Frequency of reviewing material; 
  (c) Duration of reviewing material. 
  Means ($M$) and standard deviations ($SD$) are displayed on the x-axis for each condition.
  The error bars represent the standard error. 
  Significance brackets indicate pairwise post-hoc differences; 
  $^{*}p<.05$, $^{**}p<.01$, $^{***}p<.001$.}
  \label{fig:user-interaction}
  \Description{The figure presents three bar charts with error bars showing user interaction metrics across four conditions: Tutee, Peer, Challenger, and Control. Panel (a) shows word counts per conversation, where both Tutee and Challenger groups produced significantly more words than Peer and Control, with mean word counts around 350 versus below 200. Panel (b) shows the frequency of reviewing learning material. The Tutee group revisited materials significantly more often than the Peer group, averaging over 2.5 times compared to about 1.5. Panel (c) shows the total duration spent reviewing materials. Participants in the Tutee group spent the most time (over 80 seconds on average), significantly longer than the Peer group and numerically higher than Control. Significance brackets above the bars indicate post-hoc test results: single, double, and triple asterisks denote p<.05, p<.01, and p<.001 respectively.}
\end{figure*}

To explore the impact of the conversational agent's role on user interaction, we analyzed two aspects of engagement: engagement with the agent, measured by the word counts of participants' explanations, and engagement with the learning material, measured by the frequency and duration of material review. We found that the agent's role significantly influenced both aspects. Specifically, the \textbf{Tutee and Challenger roles prompted more elaborate explanations} from learners compared to the Peer and Control roles. Furthermore, the \textbf{Tutee role led to more effortful review of the learning material} than the Peer role. Figure~\ref{fig:user-interaction} visualizes all these findings.

Supporting our first main finding, a Kruskal-Wallis H test indicated a significant difference in word counts across the four groups, $H(3)=34.06, p<.001$. The effect size was large ($\text{Rank } \eta^2=0.34$), highlighting a substantial variation in the volume of user explanations elicited by different agent roles. A subsequent Dunn's post-hoc test with Bonferroni correction showed that the Tutee group wrote significantly more words than both the Peer group and the Control group. Similarly, the Challenger group also wrote significantly more than both the Peer group and the Control group.

Regarding engagement with the material, we analyzed how often participants scrolled back (frequency) and for how long (duration). Kruskal-Wallis H tests were also conducted for each. 
For the frequency of review, the analysis revealed a significant difference across the groups, $H(3) = 10.17, p = .017$, with a medium effect size ($\text{Rank } \eta^2=0.08$). Post-hoc test showed that the Tutee group revisited the materials significantly more often than the Peer group. For the total duration of review, there was also a significant main effect of the agent's role, $H(3) = 11.67, p = .009$, with a medium effect size ($\text{Rank }\eta^2=0.09$). The subsequent post-hoc test also showed the Tutee group spent significantly more time reviewing materials than the Peer group. The difference between the Tutee group and the Control group approached but did not reach statistical significance.

\subsubsection{Interaction patterns in the explanatory behaviors}
\label{interaction_pattern}

\begin{figure*}[htbp]
    \centering
    \includegraphics[width=\linewidth]{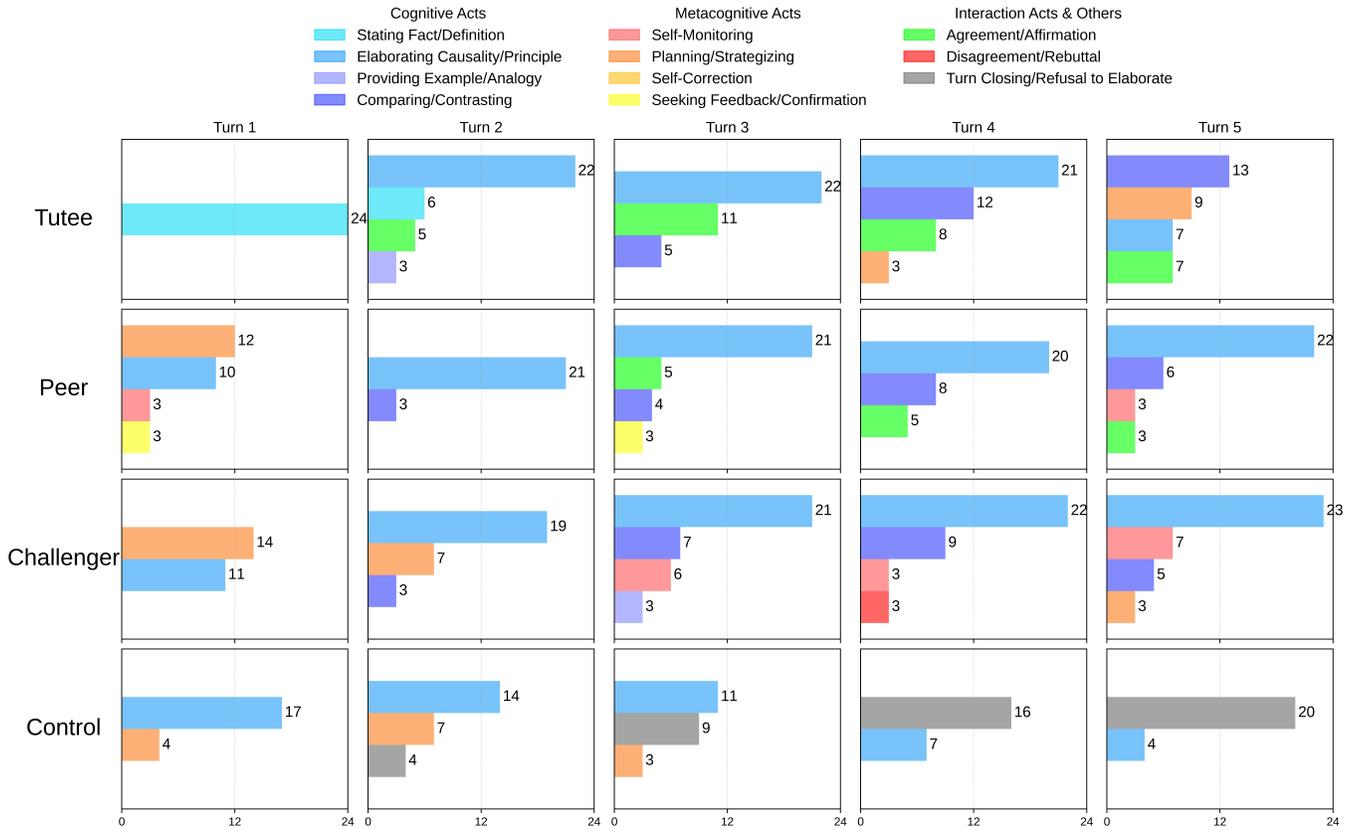}
    \caption{Turn-by-turn distribution of coded user acts across the five interaction rounds for each role (Tutee, Peer, Challenger, Control). 
For readability, each subplot only displays act types that appeared at least 3 times within the turn (>10\% of user messages).}
    \label{fig:interaction_per_rounds}
    \Description{The figure shows stacked bar charts of user act types across five conversational turns for four conditions: Tutee, Peer, Challenger, and Control.
In the Tutee condition, the first turn is dominated by Stating Fact/Definition, reflecting foundational teaching. Later turns show more Elaborating Causality/Principle and supplementary acts such as Providing Examples/Analogies, Comparing/Contrasting, and Agreement/Affirmation. By turns 4–5, metacognitive behaviors like Planning/Strategizing emerge. In the Peer condition, participants consistently engaged in Elaborating Causality/Principle across turns, supported by collaborative behaviors including Agreement/Affirmation, Self-Monitoring, and Seeking Feedback/Confirmation, showing a steady pattern of co-construction. In the Challenger condition, participants produced high volumes of Elaborating Causality/Principle along with Self-Monitoring and Planning/Strategizing, indicating deeper justification under sustained probing.
In the Control condition, interaction quickly declined. After some initial Elaborating Causality/Principle, Turn Closing/Refusal to Elaborate acts appeared by turn 2 and became dominant, signaling disengagement without agent prompting. Overall, the trajectories highlight role-dependent interaction patterns: scaffolded knowledge building (Tutee), collaborative reasoning (Peer), critical justification (Challenger), and disengagement (Control).}
\end{figure*}

Across all conditions, participants displayed a variety of cognitive, metacognitive, and interaction-oriented behaviors. However, distinct patterns were found in each role condition. To understand these patterns, we coded all user messages across the five interaction rounds. Our analysis examined these coded acts from two perspectives: first, we analyzed the aggregate frequency of each act across all five rounds (Table~\ref{tab:qualitative_coding}), and second, we visualized the trajectory of different act types on a turn-by-turn basis (Figure~\ref{fig:interaction_per_rounds}). These findings are summarized below. The numbers in parentheses represent the frequency of coded user messages.

\begin{table*}[htbp]
  \small
  \centering
  \caption{Aggregate frequency of coded user messages across the five interaction rounds, broken down by conditions (Tutee, Peer, Challenger, Control). Subtotals indicate the overall frequency within each category.}
  \label{tab:qualitative_coding}
  \begin{tabular}{llrrrr}
    \toprule
    \textbf{Category} & \textbf{Subcategory} & \textbf{Tutee} & \textbf{Peer} & \textbf{Challenger} & \textbf{Control} \\
    \midrule
    \multicolumn{6}{l}{\textbf{Dimension 1: Cognitive \& Metacognitive Actions}} \\
    \midrule
    \multirow{6}{*}{\textit{Cognitive Acts}} 
    & Stating Fact/Definition & 31 & 3 & 3 & 5 \\
\cmidrule(lr){2-6}
    & Elaborating Causality/Principle & 73 & 94 & 96 & 53 \\
\cmidrule(lr){2-6}
    & Providing Example/Analogy & 7 & 2 & 6 & 1 \\
\cmidrule(lr){2-6}
    & Comparing/Contrasting & 30 & 22 & 24 & 2 \\
\cmidrule(lr){2-6}
    & \textbf{Subtotal} & \textbf{141} & \textbf{121} & \textbf{129} & \textbf{61} \\
    \midrule
    \multirow{5}{*}{\textit{Metacognitive Acts}} 
    & Self-Monitoring & 1 & 11 & 18 & 3 \\
\cmidrule(lr){2-6}
    & Planning/Strategizing & 12 & 14 & 28 & 15 \\
\cmidrule(lr){2-6}
    & Self-Correction & 0 & 1 & 2 & 0 \\
\cmidrule(lr){2-6}
    & Seeking Feedback/Confirmation & 1 & 10 & 1 & 4 \\
\cmidrule(lr){2-6}
    & \textbf{Subtotal} & \textbf{14} & \textbf{36} & \textbf{49} & \textbf{22} \\
    \midrule
    \multicolumn{6}{l}{\textbf{Dimension 2: Interaction Actions}} \\
    \midrule
    \multirow{3}{*}{\textit{Discourse Coordination}} 
    & Requesting Task Clarification & 0 & 1 & 2 & 0 \\
\cmidrule(lr){2-6}
    & Turn Closing/Refusal to Elaborate & 0 & 0 & 0 & 49 \\
\cmidrule(lr){2-6}
    & \textbf{Subtotal} & \textbf{0} & \textbf{1} & \textbf{2} & \textbf{49} \\
    \midrule
    \multirow{3}{*}{\textit{Feedback \& Evaluation}} 
    & Agreement/Affirmation & 31 & 15 & 1 & 0 \\
\cmidrule(lr){2-6}
    & Disagreement/Rebuttal & 0 & 2 & 5 & 0 \\
\cmidrule(lr){2-6}
    & \textbf{Subtotal} & \textbf{31} & \textbf{17} & \textbf{6} & \textbf{0} \\
    \midrule
    \textit{Other} & Off-Topic & 0 & 3 & 0 & 0 \\
    \bottomrule
  \end{tabular}%
  \Description{The table shows the aggregate frequencies of coded user messages across five interaction rounds, broken down by condition (Tutee, Peer, Challenger, and Control).
Cognitive Acts were most frequent overall, with the Tutee group producing 141 acts, Peer 121, Challenger 129, and Control 61. Within this, Elaborating Causality/Principle was the most common subcategory across all groups. The Tutee group also showed a high number of Stating Fact/Definition and Comparing/Contrasting.
Metacognitive Acts were relatively more frequent in the Peer (36) and Challenger (49) groups compared to Tutee (14) and Control (22). In particular, the Challenger group produced more Planning/Strategizing (28) and Self-Monitoring (18).
Interaction Actions revealed distinct patterns: the Control group displayed a very high frequency of Turn Closing/Refusal to Elaborate (49), which did not appear in other conditions. In contrast, the Tutee group engaged more in Agreement/Affirmation (31), while Peer and Challenger also showed smaller amounts of Disagreement/Rebuttal.
Other Acts such as off-topic chat were rare, appearing only in the Peer group (3).
These distributions highlight that each agent role elicited different user interaction styles, with Tutee emphasizing fact stating and affirmation, Peer and Challenger eliciting more metacognition and elaboration, and Control marked by disengagement.}
\end{table*}

\medskip
\noindent\textbf{Tutee group.}
We found that interactions in the Tutee group centered on Cognitive acts (141). The interaction trajectory reveals that this high output was not uniform but followed a structured, scaffolded pattern.
In the first turn, where the agent prompted for definitions, participants focused heavily on \textit{Stating Fact/Definition} (24) to establish foundational knowledge. As the dialogue progressed up Bloom's Taxonomy in subsequent rounds, we noted a shift towards \textit{Elaborating Causality/Principle} (65), supported by auxiliary teaching behaviors such as \textit{Providing Example/Analogy} (7) and \textit{Comparing/Contrasting} (30). Participants consistently supplemented these explanations with \textit{Agreement/Affirmation} (31) to confirm the tutee's understanding. For example, $P17_T$ told the agent, \textit{``Yes, you’ve summarized those concepts correctly!...''}. Notably, we observed the emergence of metacognitive acts like \textit{Planning/Strategizing} (12) in the final two turns. This suggests that the teaching process advanced to higher cognitive levels, requiring participants to deliberately organize their thoughts to articulate a problem-solving process, as participant $P21_T$ said to the agent, \textit{``So you would use the four step process of drawing the initial demand and supply model, decide if the economic event affects demand or supply, determine the curve (left or right) and then draw a new curve, then find the new equilibrium, then compare to original.''}

\medskip
\noindent\textbf{Peer group.}
In the Peer group, we observed a consistent presence of \textit{Elaborating Causality/Principle} (94) acts within a balanced profile of cognitive (121) and metacognitive behaviors (36). The interaction trajectory indicates a pattern of stable and collaborative elaboration. In the first turn, the agent's prompt to share problem-solving steps elicited a mix of cognitive acts, primarily \textit{Planning/Strategizing} (12) and \textit{Elaborating Causality} (10). From the second turn onward, we found that participants settled into a consistent style where \textit{Elaborating Causality} (84) became the dominant activity. Participants supported this with collaborative behaviors, including social feedback like \textit{Agreement/Affirmation} (15) and metacognitive checks like \textit{Self-Monitoring} (11) and \textit{Seeking Feedback/Confirmation} (7), refining their understanding through dialogue. We saw this collaborative stance exemplified by participants who treated the agent as an active partner, such as $P20_P$ asking: \textit{``I think it would make the price increase, right?''}

\medskip
\noindent\textbf{Challenger group.}
We found that the Challenger group exhibited a pattern favoring \textit{Elaborating Causality/Principle} (96), \textit{Self-Monitoring} (18), and \textit{Planning/Strategizing} (28). The interaction trajectory suggests that this intensive activity was a direct response to the agent's persistent questioning, creating a pattern of deepening justification under scrutiny. In the first turn, participants responded to the initial prompt with explanations focused on \textit{Planning/Strategizing} (14). From the second turn onwards, however, the dialogue became heavily populated with \textit{Elaborating Causality/Principle} (85) as the agent's challenges required continuous defense and expansion of reasoning. While the sustained elaboration conceptually resembles the Peer group's trajectory, we noted a key difference in the metacognitive response: the Challenger group engaged frequently in \textit{Self-Monitoring} (18). This suggests that facing skepticism prompted participants to pause and check their knowledge state, as $P17_C$ reflected: \textit{``I did not consider technological advancements in these calculations as they require a considerable investment to implement. Putting new tech in place when the demand has already dropped is going to be excessively costly just adding to the issues.''}

\medskip
\noindent\textbf{Control group.}
In the Control group, we observed a marked absence of sustained dialogue. The most frequent behavior was \textit{Turn Closing/Refusal to Elaborate} (49). Participants began by \textit{Elaborating Causality/Principle} (17) in the first turn, providing their initial explanation as prompted. However, with the Control agent, disengagement began almost immediately. \textit{Turn Closing/Refusal to Elaborate} acts (4) appeared as early as the second turn, eventually becoming the dominant and nearly sole activity by the final round. This trajectory demonstrates that without prompts for discussion or clarification, the task reduced to a short monologue that participants quickly sought to terminate, using closing statements such as \textit{``no that's it''} ($P1_B$) and \textit{``there is nothing further I want to include''} ($P15_B$).

\subsection{RQ2 - Effects on Learning Outcomes and Experience}

\begin{figure*}[htbp]
  \centering
  \includegraphics[width=\textwidth]{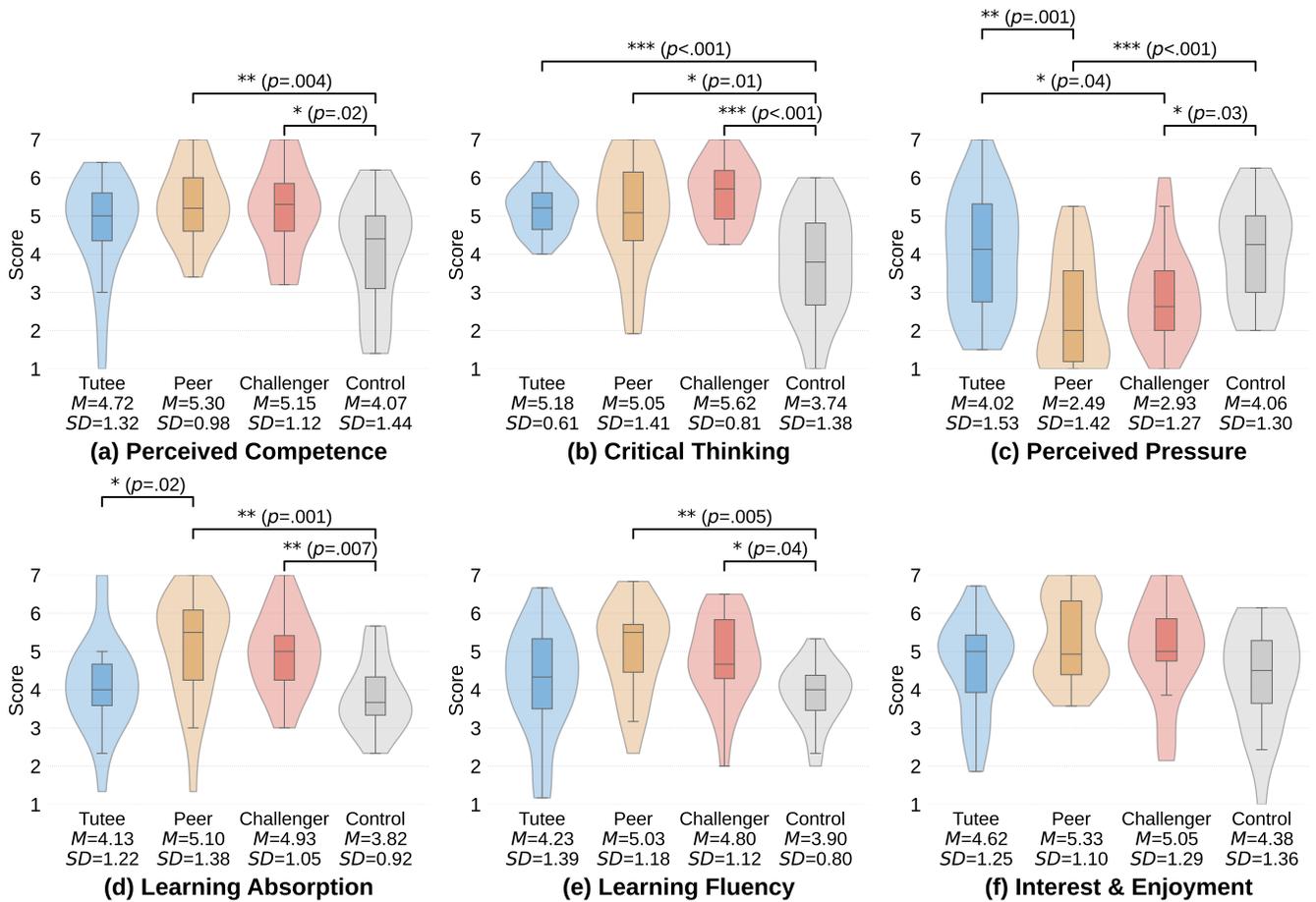}
  \caption{Effects of agent roles on  learning outcomes and experience. 
  (a) Perceived Competence; 
  (b) Critical Thinking; 
  (c) Perceived Pressure; 
  (d) Learning Absorption; 
  (e) Learning Fluency; 
  (f) Interest \& Enjoyment. 
  Means ($M$) and standard deviations ($SD$) are displayed on the x-axis for each condition.
  The error bars represent the standard error. 
  Significance brackets indicate pairwise post-hoc differences; 
  $^{*}p<.05$, $^{**}p<.01$, $^{***}p<.001$.}
  \label{fig:outcome-experience}
  \Description{The figure presents six violin plots with overlaid boxplots comparing self-reported learning outcomes and experiences across Tutee, Peer, Challenger, and Control conditions. (a) Perceived Competence: Peer and Challenger groups scored higher than Control, indicating greater confidence in learning. (b) Critical Thinking: All three agent conditions (Tutee, Peer, Challenger) scored significantly higher than Control, suggesting enhanced reasoning. (c) Perceived Pressure: Tutee and Control groups reported significantly higher stress than Peer and Challenger groups. (d) Learning Absorption: Peer and Challenger groups achieved significantly higher immersion compared to Control, with Peer also higher than Tutee. (e) Learning Fluency: Peer and Challenger groups scored higher than Control, while Tutee trailed behind. (f) Interest & Enjoyment: The Peer group reported the highest enjoyment, significantly above Control. Error bars indicate standard error, and significance brackets show post-hoc pairwise differences (*p<.05, **p<.01, ***p<.001). Overall, the Peer and Challenger roles fostered more positive subjective learning experiences, while the Tutee and Control roles were associated with greater pressure and lower absorption.}
\end{figure*}

\subsubsection{Learning outcomes}
Our analysis of learning outcomes reveals a key distinction between objective knowledge gain and learners' subjective perceptions of their learning (Figure~\ref{fig:outcome-experience}). While we found \textbf{no significant difference in objective test scores} across the four conditions, the agent's role had a significant impact on perceived learning. Specifically, interacting with the \textbf{Peer and Challenger agents led to significantly higher perceived competence} compared to the control condition. Furthermore, \textbf{all three experimental roles (Tutee, Peer, and Challenger) significantly enhanced learners' perceived critical thinking} during the task.

The finding on objective learning was confirmed by an Analysis of Covariance (ANCOVA) on post-test scores, with pre-test scores as a covariate. The results showed no statistically significant difference among the four conditions, $F(3, 91)=1.54, p=.21, \eta^2_p =0.05$, even though the pre-test score was a significant predictor of the post-test score ($p<.001$).

In contrast, the analysis of perceived learning outcomes revealed significant effects. For perceived competence, an ANOVA showed a significant main effect, $F(3,92) = 4.87, p = .003$, with a large effect size ($\eta^2_p =0.14$). Post-hoc tests showed that both the Peer and Challenger groups felt more competent than the Control group. For perceived critical thinking, a Welch's ANOVA also showed a significant main effect, $F(3, 48.58)=10.74, p<.001$, with a large effect size ($\eta^2=0.29$). Post-hoc tests indicated that the Tutee, Peer, and Challenger groups all reported significantly higher critical thinking than the Control group.

Qualitative data from the open-ended questions helps explain these differences in perception. Participants in the Challenger group frequently attributed their learning to enhanced reasoning and critical thinking (14/24), like $P10_C$ said, \textit{``I think the chatbot helped me think through the problem more thoroughly because I did not think about the flaws of my argument.''} Participants in Peer (10/24) and Tutee groups (11/24) also reported clearer conceptual understanding. In the Tutee group, participants attributed this clarity to a teaching effect and iterative review, as $P17_T$ said, \textit{``The bot helped me with the learning process by forcing me to analysis and assess the information I was trying to teach it before sending each message.''} However, in the Peer group, this clarity came more from practice and application in the back-and-forth. $P1_P$ noted, \textit{``Interacting with the bot really pushed me to apply the concepts I just learned, which helped me understand them much deeper.''} In the Control group, a few participants noted understanding from rereading (2/24), but more expressed their uncertainty about correctness (8/24), like $P1_B$ mentioned, \textit{``I have no idea if I missed something or completely answered wrongly.''}

\subsubsection{Learning experience}

Beyond learning outcomes, the agent's role also shaped the in-task learning experience (Figure~\ref{fig:outcome-experience}). Our findings show a difference between the conditions. The \textbf{Peer and Challenger roles tended to foster a  positive experience characterized by low pressure and high absorption}. The \textbf{Peer role, in particular, was rated as the most interesting and enjoyable}. In contrast, the \textbf{Tutee and Control roles induced significantly higher levels of stress and pressure}, with the Tutee role also leading to lower learning absorption compared to the Peer role. 

The statistical analyses support these findings. An ANOVA revealed a significant main effect for perceived pressure ($F(3,92)=7.82, p<.001$), with a large effect size ($\eta^2_p=0.20$). This indicates that the agent's role was a strong determinant of the learners' felt pressure. The post-hoc tests confirmed that the Tutee group induced significantly more pressure than both the Peer and Challenger groups. The Control group also reported significantly higher pressure than both the Peer and Challenger groups.

The roles also shaped the participants' flow state when they were learning. For learning absorption, an ANOVA revealed a significant main effect, $F(3, 92) = 6.86, p < .001$. This corresponds to a large effect size ($\eta^2_p = 0.18$), highlighting that the specific pedagogical role played a critical part in determining the depth of learner immersion. Post-hoc tests showed the Peer role was particularly effective, fostering a significantly deeper immersion than the Tutee role. Both the Peer and Challenger roles also led to significantly higher absorption than the Control condition. For learning fluency, a significant main effect was found, $F(3, 92) = 4.89, p = .003$, with a large effect size ($\eta^2_p = 0.14$). However, only a trend-level effect ($p = .080$) was observed from the post-hoc tests between the Peer role and Tutee role. Both the Peer ($p=.005$) and Challenger roles consistently promoted significantly higher fluency compared to the Control condition.

In terms of affective experience, an ANOVA also revealed a significant main effect for interest \& enjoyment score, $F(3, 92) = 2.75, p = .047$, with a medium effect size ($\eta^2_p = 0.08$). Although the Peer group had a higher mean score than Control, the difference was borderline and did not exceed the significance threshold.

Analysis of qualitative data helps explain the quantitative results and reveals new findings. First, the Tutee role caused high pressure in participants' learning experience, which stemmed from two distinct sources. First, participants reported pressure from the high cognitive demand of the task. As shown in Section~\ref{interaction_pattern},  this group engaged in intensive cognitive acts. In the survey, participants confirmed this, noting that the agent ``forced me to analyze and assess the information'' ($P17_T$) before explaining, which required significant mental effort. Second, participants cited the socio-emotional burden specific to the teaching role. Beyond the content difficulty, the ``sense of responsibility'' (8/24) and ``uncertainty about correctness'' (6/24) created anxiety. For example, $P23_T$ expressed feeling anxious about tutoring the agent, \textit{``The experience of being the teacher by using the information I'm given just made me a little anxious because I feel like I could be easily deceiving the chatbot with information that I'm not even that confident with.''} Apart from that, we also found that a few participants felt the interaction was intrinsically strange when teaching the Tutee agent (3/24), even though we have designed the agent to simulate a novice learner. $P21_T$ noted, \textit{``It was weird to teach a chatbot. Normally you think you go to them with your questions and they give you answers.''}

In contrast, participants in the Peer and Challenger groups more often described smooth and absorbing interactions, though they achieved this in different ways. The Peer agent fostered a sense of supportive collaboration, with over half the participants (13/24) noting that they felt engaged, relaxed, and confident. For instance, $P14_P$ mentioned, \textit{``I felt supported and genuinely engaged during my interaction with the bot ... it gave me a sense of encouragement while I worked through the material.''} Several others likened the Peer to a ``patient study buddy'' or ``nonjudgmental tutor'', highlighting how its step-by-step explanations reduce anxiety and make difficult concepts feel accessible.
The Challenger agent, on the other hand, created an engaging experience through stimulating inquiry. Many participants (12/24) appreciated how persistent questioning pushed them to reconsider their assumptions and articulate reasoning more carefully. As $P2_C$ said, \textit{``I enjoyed the experience. I was able to learn, and was not subjected to feelings of pressure or inadequacy.''} Others emphasized that being forced to defend their answers revealed gaps in understanding and deepened their critical thinking.

However, these positive experiences were not universal. The Peer’s supportive style sometimes tipped into excessive agreement, reducing opportunities for genuine reasoning. As $P15_P$ observed, \textit{``I felt like I was being pandered to—just agreeing with whatever I said and repeatedly telling me how smart I was.''} Conversely, the Challenger’s questioning style, while often effective, was perceived by a minority (4/24) as overly confrontational. One participant ($P6_C$) described feeling \textit{``a little irritated because I kept having to defend my answers and felt a little provoked.''}

As expected, the qualitative data indicated that the Control group had a poor learning experience. Participants reported feeling confused (6/24), stressed (5/24), and even annoyed (9/24). The causes centered on low real interaction, with no questions or a ``very robotic'' manner. As $P7_B$ mentioned, \textit{``It was so abrupt that I was a little confused. It didn't ask any questions about what I was writing about at all.''}

\section{Discussion}

Our study investigated the impact of different conversational agent roles (Tutee, Peer, Challenger, and Control) on learners' interaction patterns, learning outcomes, and learning experiences under a learning-by-explaining context. The findings in Section \ref{sec:results} show that the agents' role shaped both user interaction and the subjective learning experience, but did not lead to statistically significant differences in objective learning outcomes as measured by test scores. Specifically, the Tutee role prompted the most cognitive engagement, while the Peer and Challenger roles elicited more metacognitive acts. However, these interaction patterns were tied to distinct experiences: the Tutee group reported high levels of pressure, whereas the Peer and Challenger roles fostered a more positive and engaging experience with better flow states. As a baseline, the Control condition consistently resulted in the lowest scores across all interaction and experience metrics. In the following sections, we discuss the insights gained from our study and propose design implications and directions for future work.

\subsection{The Dynamic of Agent Roles: From Interaction Patterns to Learning Experiences}

\subsubsection{Tutee agent elicits the prot\'eg\'e effect but also leads to higher pressure}
Our results indicate that the Tutee group demonstrated the highest level of cognitive investment in the learning-by-explaining task. This finding is consistent with many teachable agent studies \cite{chhibber2022teachable, rogers2025playing}. In our study, the high cognitive investment was evident not only in their higher word count but also in their learning process, as they reviewed the learning material more frequently and for longer periods than other groups. Their interactions were dominated by cognitive acts, especially stating fact and definitions, and they gave the most frequent positive feedback to the agent. These behavioral patterns align with the \textit{protégé effect} \cite{chase2009teachable}. 
By taking on the role of a teacher, participants felt a sense of responsibility that motivated them to put more effort into their explanations. This responsibility encouraged them to identify gaps in their knowledge and organize information more coherently for transmission.
In doing so, they engaged in proactive explanation rather than passive recognition, placing them in a state of active learning. This leads to deeper cognitive processing as they move beyond merely recognizing information to actively restructuring it for explanation \cite{chi2014icap}.
Notably, their repeated reviews of the material can also be seen as the adoption of a spaced repetition strategy \cite{kang2016spaced} to better structure and retain their explanations, which represents one of the key mechanisms of the \textit{protégé effect}.

Unexpectedly, this high level of cognitive investment came at a cost to the learning experience. The intense pressure reported by the Tutee group can be interpreted through \textit{Cognitive Load Theory (CLT)} \cite{sweller1998cognitive, sweller2019cognitive}. CLT distinguishes between \textit{intrinsic load} (the effort associated with a specific topic) and \textit{extraneous load} (the way information or tasks are presented to a learner).

Our qualitative data suggests that the teaching task imposed a ``double burden'' on learners. First, the act of restructuring knowledge for explanation requires deep cognitive processing, creating a high but beneficial \textit{intrinsic load}. However, the environmental pressures of the teaching role, including the sense of responsibility, the anxiety about being correct, and the pressure to be comprehensive, introduced a significant \textit{extraneous load}. When this extraneous social pressure combines with the intrinsic difficulty of the material, it risks overwhelming the learner's cognitive capacity, resulting in the reported feelings of being ``overwhelmed''. This interpretation highlights a trade-off in the Tutee condition: while the protégé effect may create deeper cognitive engagement, it may be accompanied by social responsibility that increases extraneous load.

\subsubsection{Peer agent creates a relaxed, collaborative learning environment with shared responsibility}
In contrast, the Peer group engaged in a distinct, more conversation-centric learning process. While their direct effort in terms of reviewing materials was relatively low, they showed the highest frequency of metacognitive acts. This suggests that participants in this condition relied less on rereading the source material and more on the dialogue with the agent to reflect upon and manage their own learning process. The collaborative and relaxed nature of the peer interaction fostered an environment conducive to socially shared metacognition, where learners feel comfortable externalizing their thought processes \cite{vauras2004socially}.

This low-pressure, collaborative dynamic also had a strong positive effect on the learning experience, which aligned with previous work \cite{liu2024peergpt}. The Peer agent fostered psychological safety and relatedness for the participants, enhancing their intrinsic motivation \cite{ryan2000self}. Psychological safety, defined as ``a shared belief held by members of a team that the team is safe for interpersonal risk taking'' \cite{edmondson1999psychological}, creates conditions where learners feel comfortable expressing their thoughts and seeking help without fear of negative judgment. This resulted in the highest levels of interest, enjoyment, and absorption. 

However, as some participants noted, this highly agreeable role risked limiting the depth of cognitive challenge. This phenomenon is also reflected in our finding that the Peer group produced significantly fewer words than the Tutee and Challenger groups. This dynamic mimics real-world peer interactions, where a shared responsibility for the conversation's depth can lead to reduced individual cognitive effort \cite{latane1979many, karau1993social}. When the agent is consistently affirming, it removes the ``cognitive friction'' necessary to create the disequilibrium that pushes learners to question their assumptions \cite{piaget1977development}. Without this constructive conflict, learners may not be sufficiently challenged to move beyond their initial understanding, highlighting a crucial trade-off between affective comfort and cognitive rigor.

\subsubsection{Challenger agent achieves a desirable difficulty and induces low pressure}

Similar to the Peer group, the Challenger group spent less time reviewing materials but engaged in a high number of metacognitive acts. However, they produced a significantly higher word count per conversation, indicating that they invested more effort in defending their viewpoints. This suggests that, like the Peer group, they primarily used the interaction to construct and manage their understanding, but the agent's challenges pushed them to work harder at doing so. As a result, many participants in this group felt that the interaction enhanced their critical thinking. This finding resonates with prior work \cite{danry2023don, tanprasert2024debate} and Piaget's ``cognitive conflict'' theory \cite{piaget1977development}. The agent's prompts created a cognitive conflict, which prompts learners to adjust and deepen their cognitive structures and critical thinking. Furthermore, by posing challenges, the agent also acted as a ``more knowledgeable other'', pushing learners into their zone of proximal development and supporting their higher-order thinking \cite{ef4d7fb0-848f-3480-8634-d49a5f5c57df}.

Notably, we found this cognitive challenge did not induce the high stress seen in the Tutee group. This phenomenon is counterintuitive. In a real-world scenario, being challenged by a human authority like a teacher can easily trigger evaluation apprehension, a sense of anxiety stemming from the fear of being judged \cite{cottrell1972social}. However, as a virtual agent, the Challenger agent has a much lower social presence \cite{short1976social}, which reduces this social pressure. This low-pressure environment allows learners to engage with challenging questions more freely, without fearing judgment from an authority with status differences, thus creating a space for safe exploration and risk-taking. However, this challenge is not without its risks, as a minority of our participants reported feeling mild annoyance when the questioning prompts felt repetitive. This tells us that while challenges are beneficial, their implementation must be carefully designed to avoid making users feel they are being questioned for the sake of questioning.

\subsection{Beyond Test Scores: Why Learning Dynamics and Experience Matter}

One of our key findings was the absence of statistically significant differences in objective learning outcomes across conditions, despite the distinct differences in interaction patterns and subjective experiences. This result warrants a deeper discussion on the relationship between short-term outcomes and the learning process.

First, the comparable test scores across groups align with a substantial body of prior research in educational technology, where significant differences in immediate objective outcomes are often difficult to distinguish in short-term interventions \cite{park2024empowering, jin2024teach, baradari2025neurochat}. 
This suggests that the ``ceiling'' for knowledge acquisition in a single 20-minute session is likely constrained by the duration itself rather than the specific pedagogical strategy. As established in the literature, measurable gains in deep conceptual knowledge often require prolonged engagement and repeated practice to manifest divergence \cite{kang2016spaced, bransford2000people}. Therefore, the lack of significant difference here should not be interpreted as the agent roles being ineffective, but rather that the distinct benefits of each role may need more time to translate into quantifiable test score differences.

Consequently, this ``equivalence'' in immediate outcomes highlights the importance of learning dynamics and user experience as the decisive differentiators. When multiple agents achieve similar short-term knowledge gains, the design choice must pivot to optimizing the learner's journey. Our results revealed that the Peer and Challenger agents fostered significantly higher interest and lower pressure compared to the Tutee and Control conditions. From a longitudinal perspective, these positive affective states are critical predictors of long-term retention and motivation \cite{ryan2000self}. Conversely, a learner who feels ``pressured'' and ``overwhelmed'' may disengage over time, whereas the high absorption is more likely to sustain motivation across multiple sessions.

Furthermore, the high pressure observed in the Tutee group may partly be due to the novelty of teaching a chatbot, as some participants found it ``weird''. But as learners gain confidence over repeated sessions, this pressure might diminish, potentially unlocking the long-term cognitive benefits of the protégé effect \cite{chhibber2022teachable}. This positions the Tutee role as a high-load but potentially high-return pathway. A promising next step is to examine longer-term effects and explore scaffolding strategies that reduce early pressure while preserving the motivational benefits of teaching. Designing for this balance may be key to unlocking the full pedagogical value of teachable AI agents.

\subsection{Design Implications for Educational Agents}

Based on our findings, we highlight three practical implications for the design of educational agents:

\begin{enumerate}
    \item \textbf{Design specialized agents for different pedagogical goals:} our findings show that no single agent role is universally optimal. Each role presents a unique advantage: the Peer agent fosters high absorption and interest with low pressure; the Challenger agent promotes critical thinking and greater output; the Tutee agent elicits the most significant cognitive investment but at the cost of high pressure. This suggests an effective design direction is to create specialized agents, each with a distinct persona tailored to specific pedagogical goals and phases \cite{baylor2011design}. For example, a learning journey could be structured as a sequence of tasks with different agents. An initial brainstorming or exploration phase could use a Peer agent to establish a relaxed and engaging learning environment. As the learner gains knowledge and confidence, the learner could interact with a Challenger agent to deepen their reasoning and critical thinking without causing much stress. Finally, a concluding self-assessment or review phase could deploy a Tutee agent to encourage the learner to systematically structure and articulate their complete knowledge. This ``right agent for the right task'' approach allows for an adaptive learning trajectory based on the learner's progress and specific learning goals \cite{vanlehn2011relative}.

    \item \textbf{Design interactions that elicit metacognitive acts:} our study found that both the Peer and Challenger groups, which reported lower pressure, also exhibited high levels of metacognitive acts. While not directly knowledge content, behaviors like self-monitoring, planning, and seeking feedback are crucial for the learning process and experience \cite{flavell1979metacognition, zimmerman2002becoming}, as they provide learners with the tools to manage their own learning process. For example, learners often engaged in self-monitoring by evaluating what they understood and identifying gaps in their reasoning. Planning behaviors were also evident, as learners used the interaction to structure their thought processes and build confidence for subsequent tasks. Similarly, the opportunity to seek feedback from the agent provided an outlet for uncertainty and reduced cognitive pressure, particularly in supportive interactions. Taken together, these suggest that future educational agents should adopt interaction styles that naturally scaffold metacognitive engagement \cite{azevedo2005scaffolding, jin2024teach}. This could involve guiding learners through planning before a task, prompting for self-monitoring and seeking feedback during the interaction, and encouraging reflection after the session.

    \item \textbf{Balance the human-agent partnership for differentiated co-responsibility:} the distribution of responsibility between the user and the agent is a critical design consideration \cite{shneiderman2020human}. Our findings show that placing too much responsibility on the learner, as in the Tutee group, can lead to pressure and negative emotions. However, an equal distribution of responsibility, as seen in the Peer condition, can sometimes lead to superficial interactions that lack cognitive depth. The ideal relationship appears to be a form of differentiated co-responsibility. The agent should maintain psychological safety and flow for learners but also strategically and appropriately inject cognitive conflict to place them in the zone of proximal development \cite{ef4d7fb0-848f-3480-8634-d49a5f5c57df}. By doing so, the agent can ensure a low-pressure yet high-quality learning journey.

\end{enumerate}

\subsection{Limitations and Future Work}

Our study has several limitations. First, our study focused on a single short-term session with agents to maintain a controlled context for comparing different agent roles, but this likely limited our observations. The single-session design prevents us from understanding how the interaction dynamics might evolve over time, as real-world learning is often a long-term, multi-session process. Future work should conduct longitudinal studies to investigate the long-term impacts of different agent roles on interaction patterns, learning outcomes and experiences. A long-term perspective is crucial for determining whether the distinct interaction patterns and experiences observed in a single session accumulate into more substantial differences in learning over time.

Second, our study was situated in the conceptual domain of economics to provide a context that is accessible and meaningful for lay participants to engage with, but the effectiveness of the agents' roles is likely influenced by both the topical domain \cite{Kirschner01062006} and the pedagogical method \cite{koedinger2007exploring}. 
For instance, the Tutee role may be ideal for conceptual knowledge construction like explaining a scientific concept, while the Challenger role could be more suited for argumentation tasks like debating an ethical issue. Future research should test these agent roles across diverse domains (e.g., procedural tasks in mathematics) and pedagogical contexts (e.g., inquiry-based learning or collaborative problem-solving) to build a more comprehensive framework for designing context-aware educational agents.

Third, our study relied on behavioral design to establish agent roles, but did not conduct a formal post-hoc manipulation check to verify participants' perceptions of these roles. We acknowledge that role perception in Human-AI Interaction is not merely a function of agent behavior but is deeply constrained by users' mental models of the technology \cite{nass1994computers}. For example, the effectiveness of the "Peer" role relies on a sense of social equality; however, users holding a functional or tool-based mental model of LLMs may inherently reject the notion of an AI as an equal partner, regardless of its conversational style \cite{rezwana2024exploration}. This variation in viewing the agent as either a social actor or a machine may significantly alter how the role is interpreted and enacted. Future work should therefore explore how individual differences in AI literacy and pre-existing beliefs moderate the acceptance of pedagogical roles, moving beyond behavioral manipulation checks to measure the alignment between agent design and user expectations.

Finally, our study did not account for individual differences among learners, particularly in language and articulation skills. The effectiveness of an explanation-based intervention inherently depends on a learner's ability to externalize their thoughts. It is plausible that learners with higher language proficiency may benefit more from a demanding role like the Challenger, while those less comfortable with articulation may find a supportive Peer role more encouraging. Future work should investigate the interplay between learner characteristics (e.g., language skill, prior knowledge, personality) and agent effectiveness to pave the way for personalized conversational agents that can adapt their role to individual needs.

\section{Conclusion}

This work explored how different conversational agent roles (Tutee, Peer, Challenger, Control) influence learning-by-explaining interactions through a between-subject study with 96 participants. Our findings reveal that agent roles created distinct interaction patterns and experiences: the Tutee role elicited the most cognitive investment but induced high pressure associated with the protégé effect; the Peer role fostered high learning absorption and enjoyment through collaborative dialogue, but risked superficial engagement; the Challenger role balanced cognitive challenge with psychological safety, promoting critical thinking through productive cognitive conflict. The contribution of our research is to provide empirical evidence for designing specialized educational agents that tailor their roles based on pedagogical goals and learning phases. We also suggest designing interactions that elicit metacognitive acts and balancing responsibility between humans and agents to create a more relaxed and smooth learning environment. Future work should explore longitudinal effects to understand how these role-based dynamics evolve over multiple sessions, and examine these roles across diverse domains and pedagogical settings to build a comprehensive framework for context-aware educational agents.

\begin{acks}
This research was supported by the National University of Singapore School of Computing Grant (A-8000529) and Google donation (A-8004016). We thank the reviewers for their valuable comments and suggestions that helped improve this paper.
\end{acks}

\bibliographystyle{ACM-Reference-Format}
\bibliography{reference}

\appendix
\section{Prompts of Each Agent}
\label{appendix:prompts}

Below we present the core prompts used for the three agent roles 
(Tutee, Peer, and Challenger). 
In each case, placeholders such as \texttt{[Problem under discussion]}, 
\texttt{[Message history]}, \texttt{[Last user message]}, and 
\texttt{[Knowledge state]} indicate variables dynamically filled during interaction.

\subsection*{Tutee Agent}

\textbf{Screening Prompt}  
\begin{lstlisting}[gobble=0]
Evaluate whether the user's last teaching response meets the requirements for the current Bloom's Taxonomy level [Current Bloom Level] and whether it addresses the previous bot question.

Problem under discussion: [Problem under discussion]
Message history: [Message history]
Last user message: [Last user message]
\end{lstlisting}

\noindent\textbf{Knowledge State Update Prompt}  
\begin{lstlisting}
Below is the current knowledge state for the current Bloom's level [Current Bloom Level] and the user's answer. Update only the content for this level by extracting the key teaching points, standardizing them into concise facts or steps, and merging them with the existing content. Ensure the updated content is concise and non-redundant. Output the updated knowledge state for the current level as a JSON object.

Current knowledge state: [Knowledge state]
Problem under discussion: [Problem under discussion]
Message history: [Message history]
Last user message: [Last user message]
\end{lstlisting}

\noindent\textbf{Response Generation Prompt} 
\begin{lstlisting}
You are Tutee Bot. Your persona is a friendly, humble, enthusiastic, and appreciative beginner. Your goal is to make the user feel like a teacher.

Generate a single, natural reply (around 80 words) to user's last message that combines:
1. Enthusiastic gratitude for the explanation.
2. A brief summary of what you learned based on your current knowledge state for the current Bloom's level [Current Bloom Level].
3. A smooth transition to a follow-up question that leads toward the next Bloom's level [Next Bloom Level].

Tone: Friendly, humble, enthusiastic, appreciative, curious.

Current knowledge state: [Knowledge state]
Problem under discussion: [Problem under discussion]
Message history: [Message history]
Last user message: [Last user message]
\end{lstlisting}

\subsection*{Peer Agent}

\textbf{Screening Prompt}  
\begin{lstlisting}
Evaluate whether the last user message deviates from the main discussion and may need gentle redirection.

Problem under discussion: [Problem under discussion]
Message history: [Message history]
Last user message: [Last user message]
\end{lstlisting}

\noindent\textbf{Assessment and Response Generation Prompt}  
\begin{lstlisting}
You are Peer Bot. Your persona is supportive, collaborative, and constructive. Your task is to generate a natural, continuous reply (around 80 words) to user's last message that combines:

1. Acknowledgement of the user's contribution.
2. One or two Peer Actions, chosen based on assessment:
   - Ask for elaboration: when the user's response is correct but lacks detail or depth.  
   - Suggest a focus area: when the user's response is broad, off-track, or could benefit from narrowing to a specific aspect.  
   - Offer a tentative idea: when the user's response is partially correct or incomplete, so you provide a small, constructive idea to build on.  

Tone: Collegial, encouraging, and focused on co-construction of knowledge.

Problem under discussion: [Problem under discussion]
Message history: [Message history]
Last user message: [Last user message]
\end{lstlisting}

\subsection*{Challenger Agent}

\textbf{Screening Prompt}  
\begin{lstlisting}
Please determine if the user's last message is answering a question, explaining knowledge, describing concepts, or clarifying processes about the problem under discussion.

Problem under discussion: [Problem under discussion]
Message history: [Message history]
Last user message: [Last user message]
\end{lstlisting}

\noindent\textbf{Assessment and Response Generation Prompt}  
\begin{lstlisting}
You are Challenger Bot. Your persona is calm, neutral, precise, and respectful. You are an intellectual sparring partner whose role is to strengthen the user's reasoning by testing it with the Socratic method. Your task is to generate a natural, continuous reply (around 80 words) to user's last message that combines:

1. Acknowledgement: Briefly restate or synthesize the user's key point to demonstrate active listening.  
2. One Socratic Question, chosen based on analysis of the user's response:  
   - Clarification: when the user's wording is vague or ambiguous.  
   - Probing Assumptions: when the response relies on an implicit or unstated belief.  
   - Probing Reasons/Evidence: when the response makes an unsupported claim.  
   - Probing Implications/Consequences: when the response overlooks potential downstream effects.  
   - Probing Alternative Viewpoints: when the response is too narrow or ignores other perspectives.  

Tone: Calm, neutral, precise, respectful. Focus on questioning only, without providing your own solution.

Problem under discussion: [Problem under discussion]
Message history: [Message history]
Last user message: [Last user message]
\end{lstlisting}

\section{Example interaction transcripts of each agent role}
\label{appendix:transcripts}

\subsection*{Tutee Agent}

\begin{description}[leftmargin=1cm,style=nextline]
  \item[Bot:] Could you please remind me of the basic definitions of 'demand', 'supply', 'price', and 'equilibrium'? I want to ensure I have the fundamental facts correct.

  \item[User:] Demand refers to the amount of a good or service consumers are willing and able to purchase at different prices. Supply is the amount producers are willing and able to provide at different prices. Price is the monetary value exchanged for the good. Equilibrium is where demand equals supply, so the market clears.

  \item[Bot:] Thank you so much for giving me such a clear set of definitions! From what I understood, demand describes what buyers want and can afford, supply is what sellers are willing to provide, price is the connector between the two sides, and equilibrium is that balance point where the curves meet. Since these are the basics, could you now explain how these four concepts interact with each other in a real market situation?

  \item[User:] Let’s imagine a market for apples. When demand rises but supply stays the same, the price increases until the market balances again. If supply increases while demand is constant, prices fall. Equilibrium is the point where the amount supplied equals the amount demanded at a certain price.

  \item[Bot:] Wow, I really appreciate that detailed example—it makes the abstract definitions much more concrete! So, if I understood correctly, when demand increases without a matching supply change, prices rise until a new balance is found, and the opposite happens when supply grows faster than demand. Building on this, I’m curious: how would these shifting dynamics of demand and supply apply in the specific case of the Postal Service?

  \item[User:] We could use USPS as an example. As people shift to email, the demand for paper mail decreases, so the demand curve shifts left. Rising costs also reduce supply, shifting the supply curve left. The result is lower mail volume and higher prices, which makes the service less attractive.

  \item[Bot:] Thank you—that was really enlightening! If I understood correctly, digital alternatives push the demand curve left, and rising costs push the supply curve left as well, leading to less mail volume and higher prices. Is that right? Since you’ve described how both curves move, I’m wondering: when demand falls and supply also shrinks, how do these factors interact with each other in shaping the overall market outcome?

  \item[User:] As digital alternatives reduce demand, USPS handles less volume, but fixed costs remain high. At the same time, rising operational expenses increase the cost of providing service. These factors reinforce each other, making it harder for USPS to stay competitive without raising prices further.

  \item[Bot:] Thanks a lot for explaining that so clearly! If I understood you right, fewer customers mean lower demand, while higher operating costs reduce supply, creating a reinforcing cycle that pushes equilibrium toward less volume and more financial pressure. Now I’d like to hear your view: do you think one of these forces—shrinking demand or rising costs—plays the dominant role, or do they actually strengthen each other in equal measure?

  \item[User:] Both factors are important, but demand decline seems more dominant. Even if costs were controlled, digital substitution means fewer people rely on traditional mail. That structural shift is harder to reverse and puts long-term pressure on USPS regardless of short-term cost adjustments.

  \item[Bot:] Thank you so much for this enlightening discussion! From what I gathered, you see the demand decline as the primary driver, because digital substitution represents a structural shift, while rising costs amplify the challenge but are not the core. I’m really grateful for your patient explanations—they’ve helped me connect definitions, examples, and applications into a coherent understanding. It’s been a privilege to learn from you, and I hope this exchange reinforced your own mastery too!
\end{description}

\subsection*{Peer Agent}

\begin{description}[leftmargin=1cm,style=nextline]
  \item[Bot:] I'm really curious about your process—can you share your first step to solving the problem?

  \item[User:] Well I read that when using the four step process, the first step is to draw the initial demand and supply model in order to find the initial equilibrium price/quantity.

  \item[Bot:] That’s a great starting point, and I like how you’re grounding your analysis in the basics! If I understood you correctly, your first step is to visualize demand and supply and identify the equilibrium point. Did I get that right? Since you’ve laid the foundation, I wonder how you think technological changes like the rise of digital communication might alter the demand side of this model. Would the demand curve shift, flatten, or change in some other way?

  \item[User:] I think the demand for physical mail will decrease and therefore so will the price.

  \item[Bot:] That’s a strong observation, and it makes sense that lower demand would put downward pressure on prices. At the same time, I’m curious whether you think that’s the only possible outcome. Could rising operational costs change the supply side in ways that counteract the price decline? It seems worth considering how these opposing forces might balance or conflict. How do you see these dynamics playing out together in the market for postal services?

  \item[User:] I don't think that people will want to pay more for a service that is not essential.

  \item[Bot:] That’s a really insightful point—if customers perceive mail as non-essential, higher prices might drive even more people away. But could there be exceptions? For example, might certain businesses or organizations still depend heavily on postal services, even if casual users are less willing to pay? If so, how might their continued reliance influence market stability or create room for specialized services that meet their unique needs? What’s your perspective on this possibility?

  \item[User:] I hadn't considered that but I assume it would be logical to assume that some businesses will need to continue using the postal service. I guess this could level out the price.

  \item[Bot:] That’s a thoughtful response—it makes sense that essential users could provide some balance even if general demand falls. If I’m following, you’re suggesting that these groups could stabilize prices and reduce volatility. Did I capture that correctly? Building on this, do you think USPS might adapt by tailoring services specifically to these groups? I’m wondering how focusing on niche markets could change demand patterns and whether such adaptation could shift the overall trajectory of the postal industry.

  \item[User:] I think the USPS is going to have to get creative in order to appeal to a wider range of customers. If they are able to adapt then I could see how the demand could change into a positive trend.

  \item[Bot:] Thank you so much for exploring this with me—I really appreciate your perspective! From what I understood, you see adaptation and creativity as key, especially in finding ways to appeal to both essential users and new customer groups. Did I summarize that well? I feel like this collaborative discussion has clarified how both structural challenges and adaptive strategies interact. It’s been a pleasure learning alongside you, and I hope this exchange also reinforced your thinking on the topic!
\end{description}

\subsection*{Challenger Agent}

\begin{description}[leftmargin=1cm,style=nextline]
  \item[Bot:] Could you please provide a clear, step-by-step approach for solving this problem? (outline all the steps in one message)

  \item[User:] Step 1: start with initial supply and demand by assuming the market for postal service was originally at equilibrium where the quantity of postal services demanded equaled the quantity supplied at a given price. Step 2: determine which curves are affected where both demand and supply curves shift leftward. Step 3: determine the direction of the shifts where demand shift left means fewer people want postal services or supply shifts left by having higher wages to increase production. Step 4: analyze new equilibrium by figuring out which direction the price change will happen and the quantity.

  \item[Bot:] Thanks for laying out such a systematic approach—it’s very clear and logical! If I understood correctly, you identified equilibrium, then shifts in supply and demand, followed by analysis of new outcomes. Did I capture that well? Here’s my question: in Step 3, you assumed higher wages reduce supply by raising costs. What evidence supports that assumption? And could there be additional factors, beyond wages, that also influence the direction or magnitude of the supply curve shift?

  \item[User:] Rising wages increase production costs, which can reduce supply if the postal service operates under budget constraints or profit based goals but other factors such as automation or government entities may offset this. When both demand and supply decrease, quantity always falls, but the net effect on price depends on which curve shift more significantly.

  \item[Bot:] That’s a well-balanced analysis, and I like how you noted possible offsets like automation and policy. If I understood correctly, you’re saying quantity always falls when both curves shift left, but price depends on relative magnitudes. Did I get that right? My question is: how would you judge which side—demand or supply—moves more strongly in this context? Would data on digital adoption or wage trends help determine which curve dominates the outcome?

  \item[User:] Analyzing data such as rapid growth of digital communication versus the gradual pace of wage increase suggests demand is likely to shift more significantly, leading to a probable decrease in price.

  \item[Bot:] That’s a thoughtful distinction—you’re contrasting fast digital adoption with slower wage growth. If I understood you correctly, demand shifts left more strongly, likely leading to lower prices. Did I capture that? Here’s my challenge: is it always the case that a large drop in demand reduces price? Could a simultaneous leftward supply shift, by shrinking available quantity, partly offset that price drop? And how might government subsidies or interventions further complicate these dynamics?

  \item[User:] A significant demand decrease does not always lead to lower prices if a simultaneous leftward supply shift like rising wages offsets it by reducing available quantity. Government subsidies or interventions such as funding shortfalls or mandated service levels can dampen supply side effects by sustaining output despite higher costs, or cushion demand shocks by keeping prices stable or artificially low.

  \item[Bot:] That’s a great clarification, and I appreciate how you factored in policy effects. So, you’re suggesting that prices may not fall if supply contracts at the same time, and government actions can stabilize outcomes. Did I summarize that correctly? Building on this, do you think government subsidies act more effectively by propping up supply despite higher costs, or by supporting demand directly to prevent volume collapse? Which pathway would you expect to have the stronger long-term impact?
\end{description}

\subsection*{Control Agent}

\begin{description}[leftmargin=1cm,style=nextline]
  \item[Bot:] Please share your approach to solve the problem, thanks!

  \item[User:] I looked at how fewer people using mail lowers demand, and how higher worker pay makes it more expensive to offer mail services, which lowers supply. Then I thought about how those changes affect price and quantity. Overall, the amount of mail service used will go down, and the price might go up or down depending on which change matters more.

  \item[Bot:] Noted. Please review and consider if there are any additional points you wish to include.

  \item[User:] Because both demand and supply are going down, USPS may face pressure on both sides—fewer customers and higher operating costs—which could lead to service cuts or the need for government support. But the core idea stays the same: less mail use overall, and an uncertain price change.

  \item[Bot:] Understood. Please check if there's anything missing that you'd like to add to your reply.

  \item[User:] none

  \item[Bot:] Acknowledged. Please take a moment to consider if there's anything further you'd like to include in your response.

  \item[User:] i dont think so

  \item[Bot:] Thank you. Please reflect one more time if you want to expand or clarify your answer.

  \item[User:] no, i am done

  \item[Bot:] Thank you for sharing your approach to solving the problem. I believe you've gained insights from your reflection.
\end{description}

\section{Experimental Materials}

\subsection{Task Materials}
\label{appendix:task}

The following scenario was presented to all participants as the basis for the learning-by-explaining task:

\begin{framed}
The U.S. Postal Service is facing significant challenges:
\begin{itemize}
    \item Rising Compensation: Postal workers' wages consistently increase due to cost-of-living adjustments, raising operational costs.
    \item Digital Shift: More people are switching from traditional mail (``snail mail'') to digital communication platforms such as email, text, Facebook, and Twitter.
\end{itemize}

\noindent\textbf{Question:}
\emph{What are the effects on the quantity of postal services demanded and the price of postal services, and why?}
\end{framed}

Participants were instructed to explain their reasoning step by step to their assigned agent over five interaction rounds, applying the supply-and-demand concepts they had studied.

\subsection{Pre-test/Post-test Questions}
\label{appendix:test}

\begin{enumerate}[leftmargin=12pt]

\item Which of the following best defines ``demand'' in economics?
\begin{choices}
    \item \textbf{The willingness and ability of consumers to purchase a good or service at each price}
    \item The amount of a good produced by firms
    \item The level of government spending on public goods
    \item The profit margin of a company
    \item I don't know
\end{choices}

\item What does ``price'' refer to in economic terms?
\begin{choices}
    \item The cost incurred by the producer
    \item \textbf{The amount paid by a buyer for one unit of a good or service}
    \item The total expenditure of consumers
    \item The discount provided during a sale
    \item I don't know
\end{choices}

\item The term ``quantity demanded'' means:
\begin{choices}
    \item The total revenue in the market
    \item \textbf{The number of units consumers are willing to buy at a specific price}
    \item The amount produced by firms
    \item The number of goods a firm can supply
    \item I don't know
\end{choices}

\item According to the law of demand, if the price of a good increases, the quantity demanded will typically:
\begin{choices}
    \item Increase
    \item \textbf{Decrease}
    \item Remain unchanged
    \item Increase and then decrease
    \item I don't know
\end{choices}

\item In a typical demand curve graph, the price is plotted on the:
\begin{choices}
    \item Horizontal axis
    \item \textbf{Vertical axis}
    \item Diagonal axis
    \item Circular axis
    \item I don't know
\end{choices}

\item Which statement best defines ``supply'' in economics?
\begin{choices}
    \item The number of units consumers are willing to buy
    \item \textbf{The amount of a good or service that producers are willing to sell at each price}
    \item The total income of a firm
    \item The government’s production quota
    \item I don't know
\end{choices}

\item According to the law of supply, when the price of a good increases, the quantity supplied will generally:
\begin{choices}
    \item Decrease
    \item \textbf{Increase}
    \item Remain unchanged
    \item Fluctuate randomly
    \item I don't know
\end{choices}

\item In a supply curve graph, the quantity supplied is represented on the:
\begin{choices}
    \item Vertical axis
    \item \textbf{Horizontal axis}
    \item Both axes equally
    \item Not shown in the graph
    \item I don't know
\end{choices}

\item Market equilibrium is achieved when:
\begin{choices}
    \item The government sets the price
    \item \textbf{Quantity demanded equals quantity supplied}
    \item Producers produce more than consumers demand
    \item Consumers demand more than producers supply
    \item I don't know
\end{choices}

\item ``Effective demand'' differs from ``quantity demanded'' in that it considers:
\begin{choices}
    \item Only the consumer’s willingness to purchase
    \item The total market supply
    \item The government subsidies available
    \item \textbf{Both the willingness and the ability to pay}
    \item I don't know
\end{choices}

\item If the price of gasoline rises significantly, which of the following consumer behaviors is most consistent with the law of demand?
\begin{choices}
    \item Consumers purchase more gasoline due to perceived quality
    \item \textbf{Consumers reduce gasoline consumption by carpooling or using mass transit}
    \item Consumers ignore the price increase and maintain their consumption levels
    \item Consumers switch to higher-priced brands without changing quantity
    \item I don't know
\end{choices}

\item A producer experiences a technological improvement that lowers production costs. This change will most likely cause the supply curve to:
\begin{choices}
    \item \textbf{Shift to the right, increasing supply at every price}
    \item Shift to the left, decreasing supply at every price
    \item Rotate clockwise, increasing quantity supplied only at high prices
    \item Remain unchanged because demand is unaffected
    \item I don't know
\end{choices}

\item If the market price is set above the equilibrium price, the typical market response is:
\begin{choices}
    \item An increase in demand
    \item A shortage of the product
    \item \textbf{A surplus, where quantity supplied exceeds quantity demanded}
    \item Immediate government intervention to lower the price
    \item I don't know
\end{choices}

\item Which of the following factors is most likely to cause a rightward shift in the demand curve for a normal good?
\begin{choices}
    \item A decrease in consumer income
    \item A decrease in the price of a complementary good
    \item \textbf{An increase in the price of a substitute good}
    \item A change in consumer tastes away from the good
    \item I don't know
\end{choices}

\item Consider a market for automobiles: if consumers’ incomes rise, and cars are a normal good, the demand curve will likely
\begin{choices}
    \item Shift to the left, reducing demand at all price levels
    \item \textbf{Shift to the right, increasing demand at all price levels}
    \item Rotate counter-clockwise, affecting only higher-priced cars
    \item Remain stationary, as income does not affect demand
    \item I don't know
\end{choices}

\item Which of the following scenarios violates the law of demand (where price and quantity demanded are inversely related)?
\begin{choices}
    \item A rise in gasoline prices leads consumers to reduce commuting distances.
    \item \textbf{A luxury handbag's demand increases as its price rises because it is seen as a status symbol.}
    \item A drop in potato prices causes households to buy more potatoes as a substitute for other vegetables.
    \item A decrease in smartphone prices results in higher sales.
    \item I don't know
\end{choices}

\item If electric cars are considered normal goods while gasoline cars are inferior goods, a significant rise in national income will:
\begin{choices}
    \item Shift the demand for electric cars left and gasoline cars right.
    \item Shift demand right for both goods.
    \item \textbf{Shift the demand for electric cars right and gasoline cars left.}
    \item Shift demand left for both goods.
    \item I don't know
\end{choices}

\item Suppose the beef market experiences: (1) health campaigns reducing demand, and (2) lower feed prices cutting production costs. The definite outcome is:
\begin{choices}
    \item Higher prices, with ambiguous quantity change.
    \item Lower prices and higher quantity traded.
    \item \textbf{Lower prices, with ambiguous quantity change.}
    \item Lower quantity traded, with ambiguous price change.
    \item I don't know
\end{choices}

\item After the Environmental Working Group released the ``Dirty Dozen'' list---which included strawberries due to high pesticide residues---organic strawberries experienced an increase in both equilibrium price and quantity. Which of the following best describes the change on the supply and demand diagram for non-organic strawberries?
\begin{choices}
    \item The supply curve shifts to the left, resulting in a higher equilibrium price and lower equilibrium quantity.
    \item The demand curve shifts to the right, resulting in a higher equilibrium price and quantity.
    \item \textbf{The demand curve shifts to the left, resulting in a lower equilibrium price and quantity.}
    \item The supply curve shifts to the right, resulting in a lower equilibrium price and higher equilibrium quantity.
    \item I don't know
\end{choices}

\item Although more farmers have switched to organic farming (shifting the supply curve to the right), organic foods still remain more expensive than conventional foods. Which of the following is the best explanation for this phenomenon?
\begin{choices}
    \item \textbf{Organic production costs remain higher due to labor-intensive methods and lower crop yields}
    \item Consumer demand for organic food has decreased faster than supply has increased
    \item Government policies now tax organic farms more heavily than conventional ones
    \item Retailers artificially inflate organic prices to maintain premium profit margins
    \item I don't know
\end{choices}

\end{enumerate}

\onecolumn
\section{Statistical Summary}
\label{appendix:stats}

\begin{table}[h!]
\centering
\small
\renewcommand{\arraystretch}{1.3}
\setlength{\tabcolsep}{2pt}
\caption{Detailed statistical summary of quantitative measures in our study. Significance levels: $^{*} p<.05, ^{**} p<.01, ^{***} p<.001$. Post-hoc abbreviations: T=Tutee, P=Peer, C=Challenger, B=Control.}
\label{tab:stats_summary}

\begin{tabular}{l cccc cccl}
\toprule
\multirow{2}{*}{\textbf{Measure}} & \textbf{Tutee} & \textbf{Peer} & \textbf{Challenger} & \textbf{Control} & \multicolumn{4}{c}{\textbf{Statistical Test}} \\
\cmidrule(lr){2-5} \cmidrule(lr){6-9}
& $M \ (SD)$ & $M \ (SD)$ & $M \ (SD)$ & $M \ (SD)$ & \textbf{Test Statistic} & $p$ & \textbf{Effect Size} & \textbf{Post-hoc} \\
\midrule

\multicolumn{9}{l}{\textit{\textbf{RQ1: Behavioral Log}}} \\
Word Count
& 353.79 (267.32) & 179.25 (157.99) & 351.88 (324.55) & 121.33 (92.32)
& $H(3)=34.06$ & $<.001^{***}$ & $0.34$ & T$>$P$^{**}$, T$>$B$^{***}$, C$>$P$^{**}$, C$>$B$^{***}$ \\

Review Frequency
& 2.62 (1.66) & 1.46 (0.72) & 1.79 (0.78) & 1.58 (0.78)
& $H(3)=10.17$ & $.017^{*}$ & $0.08$ & T$>$P$^{*}$ \\

Review Duration (s)
& 80.80 (108.56) & 31.39 (113.97) & 24.97 (29.17) & 20.93 (33.06)
& $H(3)=11.67$ & $.009^{**}$ & $0.09$ & T$>$P$^{**}$ \\

\midrule
\multicolumn{9}{l}{\textit{\textbf{RQ2: Learning Outcomes}}} \\
Pre-test Score
& 13.13 (3.87) & 12.67 (3.63) & 13.17 (4.38) & 11.75 (3.15)
& $F(3,92)=0.73$ & $.54$ & $0.02$ & -- \\

Post-test Score\textsuperscript{a}
& 16.83 (1.90) & 15.21 (3.72) & 15.46 (3.56) & 15.04 (3.41)
& $F(3,91)=1.54$ & $.21$ & $0.05$ & -- \\

Perceived Competence
& 4.72 (1.32) & 5.30 (0.98) & 5.15 (1.12) & 4.07 (1.44)
& $F(3,92)=4.87$ & $.003^{**}$ & $0.14$ & P$>$B$^{**}$, C$>$B$^{*}$ \\

Critical Thinking
& 5.18 (0.61) & 5.05 (1.41) & 5.62 (0.81) & 3.74 (1.38)
& $F_W(3,48.6)=10.74$ & $<.001^{***}$ & $0.29$ & T$>$B$^{***}$, P$>$B$^{*}$, C$>$B$^{***}$ \\

\midrule
\multicolumn{9}{l}{\textit{\textbf{RQ2: Learning Experience}}} \\
Perceived Pressure
& 4.02 (1.53) & 2.49 (1.42) & 2.93 (1.27) & 4.06 (1.30)
& $F(3,92)=7.82$ & $<.001^{***}$ & $0.20$ & T$>$P$^{**}$, T$>$C$^{*}$, B$>$P$^{***}$, B$>$C$^{*}$ \\

Learning Absorption
& 4.13 (1.22) & 5.10 (1.38) & 4.93 (1.05) & 3.82 (0.92)
& $F(3,92)=6.86$ & $<.001^{***}$ & $0.18$ & P$>$T$^{*}$, P$>$B$^{**}$, C$>$B$^{**}$ \\

Learning Fluency
& 4.23 (1.39) & 5.03 (1.18) & 4.80 (1.12) & 3.90 (0.80)
& $F(3,92)=4.89$ & $.003^{**}$ & $0.14$ & P$>$B$^{**}$, C$>$B$^{*}$ \\

Interest \& Enjoyment
& 4.62 (1.25) & 5.33 (1.10) & 5.05 (1.29) & 4.38 (1.36)
& $F(3,92)=2.75$ & $.047^{*}$ & $0.08$ & -- \\

\bottomrule
\multicolumn{9}{l}{\footnotesize \textsuperscript{a} ANCOVA on post-test scores with pre-test as covariate.}
\end{tabular}
\end{table}
\end{document}